  \newlength\smallfigwidth
  \newlength\tinyfigwidth
  \documentclass[aps,prb,twocolumn,floatfix,amsmath,amssymb,showpacs,showkeys]{revtex4}

\newcommand{\be}{\begin{equation}}
\newcommand{\ee}{\end{equation}}
\newcommand{\ba}{\begin{align}}
\newcommand{\ea}{\end{align}}
\newcommand{\bn}{\begin{eqnarray}}
\newcommand{\en}{\end{eqnarray}}

\usepackage{graphicx}

\begin{document}

\preprint{KSU-UFSC-Wysin et al.}

\title{Thermal vortex dynamics in thin circular ferromagnetic nanodisks}
\author{G.\ M.\ Wysin}
\email{wysin@phys.ksu.edu}
\homepage{http://www.phys.ksu.edu/personal/wysin}
\affiliation{Department of Physics, Kansas State University, Manhattan, KS 66506-2601}
\author{W.\ Figueiredo}
\affiliation{Departamento de F\' isica, Universidade Federal de Santa Catarina, 
Florian\'opolis, Santa Catarina, Brazil}
\date{August 27, 2012}
\begin{abstract}
{The dynamics of gyrotropic vortex motion in a thin circular nanodisk of soft ferromagnetic
material is considered.  The demagnetization field is calculated using two-dimensional Green's 
functions for the thin film problem and fast Fourier transforms.  At zero temperature, the 
dynamics of the Landau-Lifshitz-Gilbert equation is simulated using fourth order Runge-Kutta 
integration.  Pure vortex initial conditions at a desired position are obtained with a 
Lagrange multipliers constraint.  These methods give accurate estimates of the vortex restoring 
force constant $k_F$ and gyrotropic frequency, showing that the vortex core motion is described by 
the Thiele equation to very high precision.  At finite temperature,  the second order Heun 
algorithm is applied to the Langevin dynamical equation with thermal noise and damping.  
A spontaneous gyrotropic motion takes place without the application of an external magnetic 
field, driven only by thermal fluctuations.  The statistics of the vortex radial position 
and rotational velocity are described with Boltzmann distributions determined by $k_F$ and
by a vortex gyrotropic mass $m_G=G^2/k_F$, respectively, where $G$ is the vortex gyrovector.}

\end{abstract}
\maketitle

%-------------------------------------------------------------------
\section{Introduction: Vortex states in thin nanoparticles}
%-------------------------------------------------------------------
Vortices in nanometer-sized thin magnetic particles\cite{Usov93} have attracted a lot of attention,
due to the possibilities for application in resonators or oscillators, in detectors, as objects for
data storage.\cite{Guslienko+08}  We consider the dynamic motion of an individual vortex in a thin 
circular disk (radius $R$ and height $L\ll R$) of soft ferromagnetic material such as Permalloy-79
(Py) where vortices have been commonly studied.\cite{Cowburn+99,Schneider+00}
In disks of appropriate size, the single vortex state is very stable and of lower energy than a 
single-domain state.\cite{Raabe+00}  Especially, we study the effective force constant $k_F$ 
responsible for the restoring force on a vortex when it is displaced from the disk center, 
${\bf F} = -k_F {\bf X}$, where ${\bf X}$ is the vortex core position relative to the disk 
center.  If the vortex is initially displaced from the disk center, say, by a pulsed magnetic 
field,\cite{Park++03} it oscillates in the gyrotropic mode,\cite{Guslienko++02} at an angular 
frequency $\omega_G=k_F/G$, where ${\bf G}=G\hat{z}$ is the vortex gyrovector, pointing 
perpendicular to the plane of the disk.  The vortex gyrotropic motion has been observed,
for example, by photo emission electron microscopy using x-rays.\cite{Guslienko++06}
Not only in small disks but in many easy-plane magnetic models this type of vortex dynamics has 
been studied for its interesting gyrotropic dynamics.\cite{Volkel++91,Wysin+93}  The gyrotropic 
mode is due to the translational mode\cite{Wysin96,Ivanov+98} in a whole spectrum of internal 
vibrations of a magnetic vortex.\cite{WysinVolkel96}

The force constant estimated analytically by Guslienko \textit{et al.}\cite{Guslienko++02} 
using the two-vortices model\cite{Metlov+02} gave predictions of the gyrotropic frequencies
obtained in micromagnetic simulations, with reasonably good accord between the two.  Here, we 
discuss direct numerical calculations of $k_F$ based on static vortex energies, by using a Lagrange 
multipliers technique to secure the vortex position ${\bf X}$ at a desired location,\cite{Wysin2010} 
and map out its potential within the disk.  We apply an adapted two-dimensional (2D) micromagnetics 
approach for thin systems to calculate the demagnetization field.  As a result, the calculations 
can be directly compared with the two-vortices prediction for $k_F$.  

At the same time, we give
a corresponding study of the vortex dynamics to calculate the gyrotropic frequencies. It is
found that the \textit{static} results for $k_F$ combined with the \textit{dynamics} results for
$\omega_G$ agree with the prediction of the Thiele dynamical equation,\cite{Thiele73,Huber82}
 $\omega_G=k_F/G$, to very high precision.  Similar to Ref.\ \onlinecite{Guslienko++02}, $k_F$ 
is found to be close to linear in the aspect ratio $L/R$ if the disks are thin but not so thin 
that the vortex would be destabilized.  Our values for $k_F$ are slightly less than those in the 
two-vortices model, as the numerical relaxation of the vortex structure allows for more flexibility 
than an analytic expression.

A micromagnetics study of this system\cite{Machado+11} for finite temperature shows evidence 
for a \textit{spontaneous gyrotropic vortex motion} with a radius of a couple of nanometers, 
without the application of a magnetic field.   The spontaneous gyrotropic motion occurs even 
if the vortex is initiated at the center of the disk in the simulations.  It is clear that thermal 
fluctuations should lead to a random displacement of the vortex core away from the disk center, 
however, it is striking that the ordered gyrotropic rotation appears and even dominates over the 
thermal fluctuations. 
% sentence added after review:
Here we confirm this effect, and also find that a spin wave doublet\cite{Ivanov++10}
(of azimuthal quantum numbers $m=\pm 1$) is excited together with the gyrotropic motion.

Having at hand the force constant $k_F$,  we can analyze both the dynamics and the statistics of 
the gyrotropic motion induced by the temperature.  The study of the finite 
temperature dynamics is carried out using a magnetic Langevin equation that includes stochastic 
magnetic fields together with damping.  We discuss the solution via the second order Heun 
method\cite{Garcia+98,Nowak00} applied to magnetic systems.  
Further, we introduce a technique for estimating the location of the vortex core accurately in the 
presence of fluctuations.  Based on the behavior of $k_F$ with disk geometry, we find it possible 
to predict the RMS displacement of the vortex core in equilibrium.  By using the collective coordinate 
Hamiltonian for the vortex, as derived from the Thiele equation, it is also possible to determine the 
probability distributions for vortex radial displacement $r=|{\bf X}|$ and rotational velocity 
$V=\omega_G r$.  It is interesting to see that the velocity distribution, $f(V)$, is of the Boltzmann 
form for a particle with an effective mass given by $m_G=G^2/k_F$, which is found to depend only on 
the gyromagnetic ratio $\gamma$, the magnetic permeability of free space $\mu_0$ and the disk radius.

%-------------------------------------------------------------------
\section{Discrete model for the continuum magnet}
%-------------------------------------------------------------------
We determine the magnetic dynamics for a continuum magnetic particle, but using a 
thin-film micromagnetics approach,\cite{Huang03} defining appropriate dipoles at cells of a 
two-dimensional grid.  This is a modification of usual micromagnetics\cite{Suess06} where a 3D 
grid is used.  The particle has a thickness $L$ along the $z$-axis, and a 
circular cross-section of radius $R$.  For thin film magnets it is reasonable to make the 
assumption that the 
magnetization $\vec{M}({\bf r})$ does not depend on the coordinate $z$ through the thickness.  
This is acceptable as long as the particle is very thin.  The demagnetization field tends 
to cause $M$ to lie within the $xy$ plane in most of the sample,\cite{Gioia97} except for 
the vortex core region.  Even in the vortex core, however, one should not expect large variations 
of $M$ with $z$, due to the dominance of the ferromagnetic exchange over the dipolar interactions
through short distances.  
%
% sentences added after review:
In this situation for very thin magnets, this 2D approach has the obvious advantage of greater 
speed over 3D approaches, without sacrificing accuracy.  It is somewhat like using a single layer 
of computation cells in 3D micromagnetics, with the cell height longer than its transverse dimensions.

The energy of the original continuum system, including exchange and magnetic
field energy, can be expressed as a volume integral,
\be
{\cal H} = \int dV \left\{ A \nabla \vec{m} \cdot \nabla \vec{m}
-\mu_0\left[\vec{H}^{\rm ext}+\tfrac{1}{2}\vec{H}^M \right]\cdot\vec{M} 
\right\} .
\ee
The magnetization scaled by saturation magnetization $M_s$ is used to define the scaled 
magnetization $\vec{m}=\vec{M}/M_s$, that enters in the exchange term, where $A$ is the 
exchange stiffness (about 13 pJ/m for Permalloy).  The last term is the interaction with an 
externally generated field, $\vec{H}^{\rm ext}$.  The demagnetization energy involves the 
demagnetization field $\vec{H}^M$ that is generated by $\vec{M}$, and which is determined through
a Poisson equation involving the scalar magnetic potential $\Phi_M$,
\be
\vec{H}^M = -\vec\nabla \Phi_M, \quad -\nabla^2 \Phi_M = \rho_M \equiv -\vec\nabla \cdot \vec{M}.
\ee
This is solved formally in three dimensions using a convolution with the 3D Green's function:
\bn
\Phi_M({\bf r}) &=& \int d^3{\bf r'} ~ G_{\rm 3D}({\bf r-r'}) \rho_M({\bf r'}), \\
G_{\rm 3D}({\bf r}) &=& \frac{1}{4\pi\vert {\bf r} \vert}.
\en
However, this is reduced to an effective 2D Green's operator, appropriate for thin magnetic
film problems, reviewed below.

%------------------------------------------------------------
\subsection{The micromagnetics model} 
%------------------------------------------------------------
The micromagnetics\cite{Cervera99,Garcia03} is set up to use nanometer-scaled cells in which 
to define coordinates $\vec{m}_i$ as the averaged scaled magnetization in that cell.  The 
system is divided into cells of size $a \times a \times L$ ($a\times a$ is the cross section 
in the $xy$-plane), rather than cubical cells.  Each cell $i$ contains a magnetic moment 
of fixed magnitude $\mu=La^2 M_s$, where $M_s$ is the saturation magnetization.  The 
direction of the (assumed uniform) magnetization in a cell is a unit vector, $\hat{m}_i$, 
whose dynamics is to be found.  The cells interact with neighboring cells via the exchange 
interaction, and with all other cells, due to the demagnetization field, and also with 
any external field.   

For the square grid of cells, the exchange energy is found to be equivalent to
\be
{\cal H}_{\rm ex} = - 2AL \sum_{(i,j)} \hat{m}_i \cdot \hat{m}_j
\ee
where the sum is over nearest neighbor cell pairs.  The energy scale of exchange is
taken as the basic energy unit.   Thus it is convenient to define an effective
exchange constant acting between the cells,
\be
J = 2AL ,
\ee
and for the computations, all other energies will be measured in this unit.
In addition, the saturation magnetization is a convenient unit for magnetic 
fields as well as for $\vec{M}$.  So we define scaled fields,
\be
\tilde{H}^M \equiv \frac{\vec{H}^M}{M_s}, \quad 
\tilde{H}^{\rm ext} \equiv \frac{\vec{H}^{\rm ext}}{M_s}.
\ee
As a result of this, the magnetic field interaction 
energy terms are scaled here as follows. For the demagnetization,
\be
{\cal H}_{\rm demag} = -  \frac{Ja^2}{2\lambda_{\rm ex}^2} \sum_i \tilde{H}_i^M \cdot \hat{m}_i,
\ee
and for the energy in the external field,
\be
{\cal H}_{\rm ext} = - \frac{J a^2}{\lambda_{\rm ex}^2} \sum_i \tilde{H}_i^{\rm ext} \cdot \hat{m}_i.
\ee
These depend on the definition of the exchange length,
\be
\lambda_{\rm ex} = \sqrt{\frac{2A}{\mu_0 M_s^2}},
\ee
that gives a measure of the competition between exchange and dipolar forces.
This means that the effective 2D Hamiltonian can be written as
\bn
\label{Hmm}
{\cal H} &=& - J \left\{ \sum_{(i,j)} \hat{m}_i \cdot \hat{m}_j \right. 
\nonumber \\
&& \left. + \frac{a^2}{\lambda_{\rm ex}^2} \sum_i \left( \tilde{H}_i^{\rm ext} 
   + \tfrac{1}{2}\tilde{H}_i^M \right)\cdot \hat{m}_i \right\} .
\en

%-----------------------------------------------------------------------
\subsection{The demagnetization field $\vec{H}^M$ in a thin film}
%-----------------------------------------------------------------------
It is important to calculate the demagnetization field efficiently and accurately,
as it plays an important role in the dynamics, and is the most computational effort.
An approach for thin films, described by Huang\cite{Huang03} is used here, where we 
need effective Green's functions that act in 2D on the magnetization 
$\vec{M}(x,y)$.  This is somewhat different from that used in Refs.\ \onlinecite{Wysin2010} 
and \onlinecite{Wysin+12}, where the in-plane part of $\vec{H}^M$ was calculated by first 
estimating the magnetic charge density $\rho_M$.  Here, it is preferred to calculate 
$\vec{H}^M$ directly from the field $\vec{M}$, which has less steps, and is found to 
result in extremely precise energy conservation in the absence of damping.

By applying an integration by parts, and throwing out a surface term outside the
magnet, the solution for the magnetic potential is first written as an operation 
on $\vec{M}$:
\be
\Phi_M({\bf r}) = \int d^3{\bf r'}~ \vec\nabla' G_{\rm 3D}({\bf r-r'}) \cdot \vec{M}({\bf r'}).
\ee
One can notice that this involves the propagator for the dipole potential, that is,
\be
\vec\nabla' G_{\rm 3D}({\bf r-r'}) = \frac{{\bf r-r'}}{4\pi \vert {\bf r-r'} \vert^3}
\ee
is the function whose product with a source dipole at position ${\bf r'}$ gives the magnetic
potential at ${\bf r}$ due to that dipole.   

To proceed further, it is useful to consider the contributions to the vertical ($z$) and horizontal 
($xy$) components of $\vec{H}^M$ separately.  Consider a source cell centered at $(x',y')$, 
and the vertical component of $\vec{H}^M$ it generates, due to $M_z'\equiv M_z(x',y')$,
at an observer position $(x,y)$. The usual procedure is to \textit{sum} over the source point 
$z'$ and \textit{average} over the observer position $z$.  One has the contribution from this 
cell, of area $dA'=dx' dy'$,
\be
d\Phi_M = \frac{dA' M_z'}{4\pi} \int_{-\delta}^{\delta} dz' ~ \frac{(z-z')}{[\tilde{r}^2+(z-z')^2]^{3/2}}
\ee
where $\delta=L/2$ and the notation $\tilde{r}^2=(x-x')^2+(y-y')^2$ is used.   The integration gives
\be
d\Phi_M = \frac{-dA' M_z'}{4\pi} \left[ \frac{1}{\sqrt{\tilde{r}^2+(z+\delta)^2}}
- \frac{1}{\sqrt{\tilde{r}^2+(z-\delta)^2}} \right] .
\ee
This would also be obtained exactly the same if starting from the magnetic surface charge density.
Then, its negative gradient with respect to $z$ gives the contribution to the demagnetization field.
If we also do the averaging over the observer position $z$, these two operations undo each other.
The field averaged in the observer cell position is
\be
\langle dH_{Mz}\rangle = - \frac{1}{L} \int_{-\delta}^{\delta} dz \frac{d}{dz}\Phi_M
= - \left.\frac{1}{L} d\Phi_M \right\vert_{-\delta}^{+\delta}.
\ee
Evaluation of the limits, and then including a sum over the source point ${\bf r'}=(x',y')$, 
shows that the field is determined by convolution with an effective Green's function in 2D,
\bn
H_{Mz}({\bf r}) &=& \int d^2{\bf r'}~ G_{zz}({\bf r-r'}) M_z({\bf r'}), \\
G_{zz}({\bf \tilde{r}}) &=& \frac{1}{2\pi L}\left( 
\frac{1}{\sqrt{{\tilde{r}}^2+L^2}}-\frac{1}{\tilde{r}} \right) .
\en
In these expressions, it is understood that the positions ${\bf r}$, ${\bf r'}$ and the 
displacement between the two, ${\bf \tilde{r}}={\bf r-r'}$, are now two-dimensional. The expression
for $G_{zz}$ is divergent at zero radius.  However, it is a weak divergence that can be regularized 
for the computation on the grid, by averaging over the cell area.  For removing the divergence at 
$\tilde{r}=0$, averaging over a circle of area equal to the cell area $a^2$ replaces the value of $G_{zz}(0)$
by the longitudinal demagnetization factor $N_z$ for cylinder of length $L$ and radius $r_o=a/\sqrt{\pi}$.
So we set
\be
G_{zz}(0) = \langle G_{zz} \rangle_o = -N_z = - \frac{1}{L}\left( L+r_o-\sqrt{L^2+r_o^2} \right).
\ee 
The "o" subscript refers to averaging over the circle of radius $r_o$.  See Ref.\ 
\onlinecite{Wysin2010} for more details. Note that $G_{zz}$ is always negative; it correctly gives 
the demagnetization field opposite to the magnetization $\vec{M}$ which generated $\vec{H}^M$.

For the in-plane components of $\vec{H}^M$, a similar procedure can be followed.  
Due to symmetry considerations, only $M_x$ and $M_y$ can contribute. One can start
by finding the magnetic potential,
\be
d\Phi_M = \frac{dA'}{4\pi} \int_{-\delta}^{\delta} dz' ~ \frac{(x-x')M_x' +(y-y')M_y'}{[\tilde{r}^2+(z-z')^2]^{3/2}} .
\ee
The integration over the source vertical coordinate $z'$ gives
\bn
d\Phi_M &=& \frac{dA'}{4\pi \tilde{r}^2} \left[ (x-x') M_x' +(y-y') M_y'\right]
\nonumber \\
& \times & \left[ \frac{z+\delta}{\sqrt{\tilde{r}^2+(z+\delta)^2}}
- \frac{z-\delta}{\sqrt{\tilde{r}^2+(z-\delta)^2}} \right].
\en
The averaging over the observer point $z$ can be carried out, and gives,
\bn
\langle d\Phi_M \rangle &=& \frac{1}{L} \int_{-\delta}^{\delta} dz ~ d\Phi_M(z) 
= \frac{\sqrt{{\bf \tilde{r}}^2+L^2}-|{\bf \tilde{r}}|}{2\pi L {\bf \tilde{r}}^2} \nonumber \\
& \times & \left[ (x-x') M_x' +(y-y') M_y' \right] dA' .
\en
Finally, the in-plane gradient leads to the in-plane demagnetization components.  Including 
also the $z$ components,  the demagnetization field averaged in the observer cell is obtained from
\bn
\vec{H}^{M}_{\alpha}({\bf r}) &=& \int d^2{\bf r'}  \sum_{\beta=x,y,z}
G_{\alpha\beta}({\bf r-r'})M_{\beta}({\bf r'}) .
\en
The elements of the Green function needed here are found to be
\bn
G_{xx}({\bf \tilde{r}}) &=& \frac{L}{2\pi \tilde{r}^4} 
\left( \frac{\tilde{x}^2}{\sqrt{\tilde{r}^2+L^2}}
-\frac{\tilde{y}^2}{\sqrt{\tilde{r}^2+L^2}+\tilde{r}} \right) ,  \\
G_{xy}({\bf \tilde{r}}) &=&  \frac{L}{2\pi \tilde{r}^4} 
\frac{2\sqrt{\tilde{r}^2+L^2}+\tilde{r}}
{\sqrt{\tilde{r}^2+L^2}+\tilde{r}} \frac{\tilde{x}\tilde{y}}{\sqrt{\tilde{r}^2+L^2}} .
\en
The element $G_{yy}$ is obtained from $G_{xx}$ by swapping $x$ and $y$ indices, and $G_{yx}=G_{xy}$.
One can verify that these matrix elements go over to those for the far-field of a point dipole, in the 
limit $\tilde{r}\rightarrow \infty$.

These transverse elements of $G$ also are not defined at zero radius, because an implicit assumption in 
the derivation is that the observation point is outside of the source cell.  There needs to be an internal
demagnetization effect within a cell even for a transverse  magnetization such as $M_x\ne 0$ or $M_y\ne 0$.  
For long thin cells with $L\gg a$, this internal transverse demagnetization factor would  be approximately 
$N_x = N_y \approx \tfrac{1}{2}$.  As a better alternative, we set $G_{xy}(0)=0$, and replace $G_{xx}(0)$ and 
$G_{yy}(0)$ with the transverse demagnetization factor of a cylinder with cross-sectional radius 
$r_o=a/\sqrt{\pi}$,
\be
G_{xx}(0)=G_{yy}(0)=-N_x = \frac{1}{2L}\left( \sqrt{L^2+r_o^2}-r_o \right).
\ee
In this way, the internal demagnetization components of the computation cells  satisfy the requirement 
$N_x+N_y+N_z=1$, while making $G_{xx}(0)$ and $G_{yy}(0)$ consistent with the regularization done for $G_{zz}(0)$.

The above results show that $\vec{H}^M$ is found by convolution of the 2D Green's operator, as a matrix,
with $\vec{M}$.  The calculation can be made faster by using a fast Fourier transform (FFT) 
approach,\cite{Sasaki97} which replaces the convolution in real space with multiplication in 
reciprocal space.  Of course, the simplest FFT approach requires a grid with a size like 
$2^{n} \times 2^{n}$, where $n$ is an integer. Our 2D system is a circle of radius $R=N a$ 
($N$ is the size in integer grid units).  For the FFT approach to work, so that the system 
being simulated is a single copy of the circle with no periodic interactions with the images, one can 
choose the smallest $n$ such that $2^{n}\ge 2N$.  By making the FFT grid at least twice as large as 
the circle to be studied, the wrap-around problem, due to the periodicity of Fourier transforms, is 
avoided in the evaluation of the convolution.  The FFT of the Green's matrix, which is static, is 
done only once at the start of the calculation.  During every time step of the integrations, however, 
the FFT of the magnetization field components must be carried out, for every stage at which the 
demagnetization field is required.  Of course, the inverse FFTs to come back to $\vec{H}^M$ 
are needed as well in every stage of the time integrator.

%--------------------------------------------
\section{The dynamics and units}
%--------------------------------------------

%--------------------------------------------
\subsection{Zero temperature}
%--------------------------------------------
The zero-temperature undamped dynamics of the system is determined by a torque equation,
for each cell of the micromagnetics system, 
\be
\frac{d\vec\mu_i}{dt} = \gamma \vec\mu_i \times \vec{B}_i.
\ee
Here $\vec{B}_i$ is the local magnetic induction acting on the $i^{\rm th}$ cell, $\gamma$ is the 
electronic gyromagnetic ratio, and the dipole moment of the cell is $\vec{\mu}_i = La^2 M_s \hat{m}_i$.  
The local magnetic induction can be defined supposing an energy $-\vec\mu_i\cdot \vec{B}_i$ for each dipole,
with
\bn
\label{Bi}
\vec{B}_i &=& -\frac{\delta {\cal H}}{\delta \vec\mu_i} = -\frac{1}{\mu}\frac{\delta {\cal H}}{\delta \hat\mu_i}
= \frac{J}{La^2 M_s} \vec{b}_i, 
\nonumber \\
\vec{b}_i & \equiv & \sum_{j=z(i)} \hat{m}_j 
+ \frac{a^2}{\lambda_{\rm ex}^2} \left( \tilde{H}_i^{\rm ext}+\tilde{H}_i^M \right) .
\en
The sum over $j$ contains only sites $z(i)$ that are nearest neighbors of site $i$.  This dimensionless 
induction $\vec{b}_i$ used in the simulations is converted to real units by the following unit of 
magnetic induction, 
\be
B_0 \equiv \frac{J}{La^2 M_s} = \frac{2A}{a^2 M_s} 
= \frac{\lambda_{\rm ex}^2}{a^2}\,\mu_0 M_s .
\ee
For computations, the dynamics is written in terms of the dimensionless fields, also scaling the
time appropriately:
\be
\frac{d\hat{m}_i}{d\tau} = \hat{m}_i \times \vec{b}_i, \quad
\tau = \gamma B_0 t.
\ee
This means that the unit of time in the simulations is $t_0=(\gamma B_0)^{-1}$.
For Permalloy with $A=13$ pJ/m, $M_s=860$ kA/m,  one has $\lambda_{\rm ex}\approx 5.3$ nm.
In our simulations we put the transverse edge of the cells as $a=2.0$ nm.  Then using the 
gyromagnetic ratio, $\gamma = e/m_e \approx 1.76 \times 10^{11}$ T$^{-1}$ s$^{-1}$, the computation 
units are based on $\mu_0 M_s=1.08$ T and  $B_0\approx 7.59$ T.  This large value for $B_0$ is 
the scale of the local magnetic induction due to the exchange interaction between the cells.  
The time unit is then $t_0 \approx 0.75$ ps; a frequency unit is $f_0=\gamma B_0=1.336$ THz.  
We may display frequency results, however, in units of $\tfrac{\mu_0}{4\pi} \gamma M_s \approx $ 15.1 GHz
for Permalloy, as this expression is equivalent to $\gamma M_s$ in CGS units.  For the disk sizes used here, 
typical periods of the vortex gyrotropic motion are around $\tau_G \sim 4000$, which then corresponds 
to dimensionless frequency $\nu=1/\tau_G \sim 2.5\times 10^{-4}$, and hence, physical frequency 
$f=\nu f_0 \sim 0.3$ GHz.

In some cases we also need to include Landau-Gilbert damping, with some dimensionless
strength $\alpha$.  Then this is included into the dynamics with the usual modification,
\be
\frac{d\hat{m}_i}{d\tau} = \hat{m}_i\times \vec{b}_i 
- \alpha \hat{m}_i\times \left( \hat{m}_i\times \vec{b}_i \right).
\ee
The zero temperature dynamics was integrated numerically for this equation, using a
standard fourth-order Runge-Kutta (RK4) scheme. Typically, a time step of $\Delta\tau=0.04$
was found sufficient to insure the correct energy conserving dynamics (when $\alpha=0$) 
and result in total energy conserved to better than 12 digits of precision over $5.0\times 10^5$
time steps in a system with as many as 4000 cells.  To get this high precision, however, 
it is necessary to always evaluate the full demagnetization field at all four intermediate
stages of the individual Runge-Kutta time steps.

%-------------------------------------------------------------------
\subsection{Finite temperature: Langevin dynamics}
%-------------------------------------------------------------------
For non-zero temperature, the dynamics is investigated here using a Langevin approach.  This
requires including both a damping term and a stochastic torque in the dynamics; together
they represent the interaction with a heat bath.  The size of the stochastic torques is related 
to the temperature and the damping constant, such that the system reaches thermal equilibrium.

It is reasonable to think of the dynamics depending on \textit{stochastic magnetic inductions} $\vec{b}_s$,
in addition to the deterministic fields $\vec{b}_i$ from the Hamiltonian dynamics. For the discussion
here, suppose we consider the dynamics of one computation cell, and suppress the $i$ index.
The dynamical equation for that cell's $\hat{m}$, including both the deterministic and random fields, is
\be
\label{LV}
\frac{d{\hat{m}}}{d\tau} = \hat{m} \times \left(\vec{b}+\vec{b}_s \right)
-\alpha \hat{m} \times \left[ \hat{m} \times \left(\vec{b}+\vec{b}_s \right) \right].
\ee
The first term is the free motion and the second term is the damping. 
Alternatively, the dynamics can be viewed as that due to the superposition of the
deterministic effects (due to $\vec{b}$) and stochastic effects (due to $\vec{b}_s$).

For a given temperature $T$, the stochastic fields establish thermal equilibrium, 
provided the time correlations satisfy the fluctuation-dissipation (FD) theorem,
\be
\langle b_s^{\lambda}(\tau) \, b_s^{\lambda'}(\tau')\rangle = 2\alpha \,  
{\cal T}\, \delta_{\lambda\lambda'} \, \delta(\tau-\tau') .
\ee
$\delta_{\lambda\lambda'}$ is the Kronecker delta and the indices $\lambda,\lambda'$ 
refer to any of the Cartesian coordinates;  $\delta(\tau-\tau')$ is a Dirac delta function.
The dimensionless temperature ${\cal T}$ is the thermal energy scaled by the 
energy unit $J$,
\be
{\cal T} \equiv \frac{kT}{J} = \frac{kT}{2AL},
\ee
where $k$ is Boltzmann's constant.  The fluctuation-dissipation theorem expresses how the 
power in the thermal fluctuations is carried in the random magnetic fields.  In terms of the 
physical units, the relation is
\be
\gamma \mu \langle B_s^{\lambda}(t) B_s^{\lambda'}(t')\rangle = 2\alpha \, 
kT\, \delta_{\lambda\lambda'} \, \delta(t-t') .
\ee
where $\mu=La^2 M_s$ is the magnetic dipole moment per computation cell.

%-------------------------------------------------------------------------------------------
\subsection{Time evolution with second order Heun (H2) method} 
%-------------------------------------------------------------------------------------------
The Langevin equation (\ref{LV}) is a first-order differential equation that is
linear in multiplicative noise. If $y=y(\tau)$ represents the full state of the system 
(a vector of dimension $3N$, where $N$ is the number of cells), then the dynamics follows 
an equation of the form
\be
\label{simple}
\frac{dy}{d\tau} = f[\tau,y(\tau)] + f_s[\tau,y(\tau)] \cdot {b}_s(\tau).
\ee
The vector function $f$ is the deterministic time derivative and the vector function 
$f_s$ determines the stochastic dynamics; $b_s$ represents the whole stochastic field 
of the system.  An efficient method for integrating this type of equation forward in time is 
the second order Heun (H2) method.\cite{Garcia+98,Nowak00}  That is in the family of 
predictor-corrector schemes and is rather stable. It involves an Euler step as the predictor 
stage, and a corrector stage that is equivalent to the trapezoid rule. Some details of the 
method are summarized here, to indicate how the stochastic fields are included, and to show 
why it is used rather than the fourth order Runge-Kutta method (the latter seems difficult to 
adapt to the stochastic fields).

We use the notation $y_n\equiv y(\tau_n)$ to show the values at times $\tau_n=n\Delta\tau$,
according to the choice of some integration time step $\Delta\tau$.  Integrating Eq.\ 
(\ref{simple}) over one time step gives the Euler predictor estimate for $y(\tau_n+\Delta\tau)$: 
\be
\tilde{y}_{n+1}  = y_n + f(\tau_n,y_n) \Delta \tau + f_s(\tau_n,y_n) \cdot (\sigma_s w_n).
\ee
The last factor, $\sigma_s w_n$, is introduced to represent the time-integral of the stochastic magnetic 
inductions. $\sigma_s$ is a variance and $w_n$ represents a vector of $3N$ random numbers, one for each
Cartesian component at each site of the grid.  Consider, say, the result of integrating the equation of
motion for just one component for one site:
\be
\int_{\tau_n}^{\tau_n+\Delta \tau}  d\tau ~ b_s^x(\tau) \longrightarrow \sigma_s w_n^x.
\ee
The physical variance $\sigma_s$ needed for this to work correctly, must be determined by the FD theorem.
For this individual component at one site, the squared variance is
\bn
\sigma_s^2 &=& \left\langle \left( \int_{\tau_n}^{\tau_n+\Delta\tau} d\tau ~ b_s^x(\tau) \right)^2 \right\rangle 
\nonumber \\
&=& \int_{\tau_n}^{\tau_{n}+\Delta\tau} d\tau \int_{\tau_n}^{\tau_n+\Delta\tau} d\tau'  ~
\left\langle  b_s^x(\tau)  b_s^x(\tau') \right\rangle  .
\en
Now applying the FD theorem to this gives the required variance of the random fields, that depends
on the time step being used:
\be
\sigma_s= \sqrt{2\alpha {\cal T}\, \Delta \tau}.
\ee
This means that individual stochastic field components $b_s^{\lambda}(\tau)$, integrated over one time 
step, are replaced by random numbers of zero mean with variance $\sigma_s$, as used above.

For the corrector stage, the points $y_n$ and $\tilde{y}_{n+1}$ are used to get better 
estimates of the slope of the solution. Then their average is used in the trapezoid corrector stage:
\bn
y_{n+1} &=& y_n + \frac{1}{2}\left[ f(\tau_n,y_n) +f(\tau_{n+1},\tilde{y}_{n+1}) \right] \Delta\tau
\\
&+& \frac{1}{2} \left[ f_s(\tau_n,y_n) + f_s(\tau_{n+1},\tilde{y}_{n+1}) \right] \cdot (\sigma_s w_n).
\nonumber
\en
The error is of order ${\cal O}((\Delta\tau)^3)$, hence it is a second order scheme.
Note that the same vector of $3N$ random numbers $w_n$ used in the predictor stage are 
re-used in the corrector stage, because it is the evolution over the same time interval.

In the coding for computations, one  does not use the explicit form of the functions
$f$ and $f_s$.  Rather, at each cell, first one can calculate the deterministic effective
field $\vec{b}_i$ based on the present state of the system.  Its effect in the dynamics will 
be actually proportional to its product with the time step, i.e., it gives a contribution 
$\Delta \hat{m}_i \propto \vec{b}_i \Delta\tau$.  Of course, the stochastic change in 
this same site will be proportional to the stochastic effective field, which is some 
$\sigma_s \vec{w}_i$ for that site, where $\vec{w}_i=(w_i^x,w_i^y,w_i^z)$. So the total change 
at this site is linearly determined by a combination,
\be
\Delta \hat{m}_i \propto  \vec{g}_i, \quad 
\vec{g}_i\equiv \vec{b}_i \Delta\tau + \sigma_s \vec{w}_i.
\ee
An effective field combination $\vec{g}_i$ acts in this way both during the predictor and
the corrector stages.  In either stage, a dynamic change in a site is given by a simple relation,
\be
\Delta \hat{m}_i = \hat{m}_i \times \left[ \vec{g}_i -\alpha (\hat{m}_i\times \vec{g}_i) \right].
\ee
Of course, the predictor stage uses the last configuration of the whole system to determine
all the $\vec{b}_i$, while the corrector finds the needed $\vec{b}_i$ based on the predicted
positions.  And, the corrector actually does the average of $\Delta \hat{m}_i$ from the
Euler stage and the second estimate from the corrector stage.  The same random numbers $w_n$
used in the predictor stage are used again in the corrector, for a chosen time step.

The integration requires a long sequence of quasi-random numbers $w_n$.  It is important that 
the simulation time does not surpass the period of the random numbers.  We used the generator 
{\tt mzran13} due to Marsaglia and Zaman, \cite{mzran94} implemented in the C-language for 
long integers. This generator is very simple and fast and has a period of about $2^{125}$, 
and is based on a combination of two separate generators with periods of $2^{32}$ and $2^{95}$.

%-------------------------------------------------------------------
\section{Vortex state properties and zero-temperature dynamics}
%-------------------------------------------------------------------
The dynamics at zero temperature, calculated with RK4, was used to check basic vortex dynamic
properties such as the stability and gyrotropic mode frequency.  We also used the Langevin 
dynamics calculated with second order Heun method to include finite temperature to see the 
primary thermal effects for some specific vortex initial configurations.  For some of these
studies, it is extremely beneficial to produce a well-formed initial vortex state in some
desired location without the presence of spin waves.

An initial vortex state is prepared first in a planar configuration of positive vorticity 
$q=+1$, namely, in-plane magnetization angle $\phi=\tan^{-1}m_y/m_x$ given by
\be
\phi(x,y) = q\tan^{-1} \frac{x-x_0}{y-y_0}.
\ee
(The negative vorticity state $q=-1$ is destabilized by the demagnetization field, so
there is no reason to consider it.)
This is the profile of a vortex centered at position $(x_0,y_0)$.  The out-of-plane
component here is $m_z=0$, however, the stable vortex state has a nonzero out-of-plane
component close to $m_z=\pm 1$ at the vortex core (polarization $p=\pm 1$). This stable
vortex state was reached by the local spin alignment procedure\cite{Wysin96} for a vortex at the 
constrained position $(x_0,y_0)$, described in Ref.\ \onlinecite{Wysin2010}. Briefly,
that is a procedure where each $\hat{m}_i$ is aligned along its local induction $\vec{b}_i$,
and the process is iterated until convergence.  The constraint is applied as extra fictitious
fields included with the Lagrange multiplier technique, that force the desired vortex starting 
position.  This procedure helps to remove any spin waves that would otherwise be generated 
starting from any arbitrary  initial state.  This state would be a perfect static state if
generated in the center of the disk.  When generated off-center, the dynamics associated
with its motion still is able to produce some spin waves.  A cleaner vortex motion can be
generated if there is a weak damping applied ($\alpha=0.02$) over some initial time interval
($\tau\approx 1000$).  After that, the system can be let to evolve in energy-conserving
dynamics, if needed.

This relaxed vortex state develops either positive or negative out-of-plane component,
including some small randomness in the initial state before the relaxation.  If $m_z \approx +1$ ($-1$)
in the vortex core region, the vortex has positive (negative) polarization and a positive (negative) 
gyrovector $G=G_z$, defined from
\be
{\bf G} = 2\pi Q \frac{m_0}{\gamma} \, \hat{z}, \quad Q\equiv qp.
\ee
$\gamma$ is the electron gyromagnetic ratio and $m_0=\mu/a^2=LM_s$ is the magnetic dipole moment per unit 
area.  The integer $Q=\pm 1$ defines the quantized topological charge that determines the two allowed discrete 
values of the gyrovector.  To a good degree of precision, the vortex states studied here obey a dynamics 
for the vortex velocity ${\bf V}$ described by a Thiele equation,\cite{Thiele73,Huber82} ignoring any 
intrinsic vortex mass\cite{Wysin96} or damping effects, 
\be
{\bf F}+{\bf G}\times {\bf V} = 0.
\ee
This equation comes from an analysis of the Hamiltonian dynamics of a magnetic system,\cite{Wysin+91,Volkel++91} 
in which the vortex excitation profile preserves its shape but moves with some collective coordinate 
center position ${\bf X}(t)$, with ${\bf V}(t)=\dot{\bf X}(t)$.  The force ${\bf F}$ is the 
gradient of the potential experienced by the vortex. The force points towards the nanodisk 
center, and can be approximated by some harmonic potential with force constant $k_F$, for 
a vortex at distance $r$ from the center,
\be
{\bf F}= -k_F r \, \hat{r}.
\ee
Hence, the presence of the gyrovector leads to the well-known gyrotropic (or uniform
circular) motion. Solving for the vortex velocity results in
\be
\label{Vvort}
{\bf V} = \frac{\hat{z}\times {\bf F}}{G} = - \frac{\gamma k_F r}{2\pi Q LM_s} \hat{\phi}.
\ee
$G$ includes the sign of the gyrovector (vector ${\bf G}$ points perpendicular to the plane of
the disk, and it has only a $z$ component).  Thus, the vortices generated with positive 
(negative) gyrovector move clockwise (counterclockwise) in the $xy$ plane.  Furthermore, 
the angular frequency of this gyrotropic motion is given by a related equation,
\be
\label{wG}
\omega_G = \frac{V}{r} = -\frac{k_F}{G} = - \frac{\gamma k_F}{2\pi Q LM_s}.
\ee
The force constant has been estimated theoretically from the rigid vortex approximation\cite{Guslienko++01} 
and from the two-vortex model.\cite{Guslienko++02}  Below, we determine $k_F$ numerically from  relaxed 
vortex states\cite{Wysin2010} (a flexible vortex).  The frequency in Eq.\ (\ref{wG}) applies to the 
stable vortex states.  If the disk is too thin, the vortex could be unstable; this produces an outward 
force ${\bf F}$, and results in the gyrotropic motion in the ``wrong'' direction.  Thus it is easy to 
identify whether a vortex is stable or unstable from a short integration of its dynamics.

In the time and frequency units applied in the simulations, the dimensionless gyrotropic frequency 
$\Omega_G$ is obtained from
\be
\label{OG}
\Omega_G=\omega_G t_0 = \frac{\omega_G}{\gamma B_0}
= -\frac{k_F a^2}{4\pi L A Q}  .
\ee
The negative sign shows that vortices with a negative gyrovector ($Q=-1$) have a
counterclockwise rotational motion; the opposite sense holds for positive gyrovector.
The force constant $k_F$ increases with thickness $L$ but decreases with disk radius $R$.  
Therefore, in the simulation time units, the gyrotropic frequency could depend primarily on
their ratio, $L/R$.  

For detection of the vortex motion, one method is to measure the spatially averaged magnetization,
\be
\langle \vec{m} \rangle = \frac{1}{N} \sum_i \vec{m}_i .
\ee
This is a useful measure of vortex gyrotropic motion, especially for experiments, where
it may not be possible to observe the rapidly changing instantaneous vortex core position. However, 
$\langle \vec{m} \rangle$  can show rotational oscillations even when no vortex is present.  Thus, we 
need instead a measure of the vortex core position based on the location of the vorticity charge center.  

\begin{figure}
\includegraphics[width=\smallfigwidth,angle=-90]{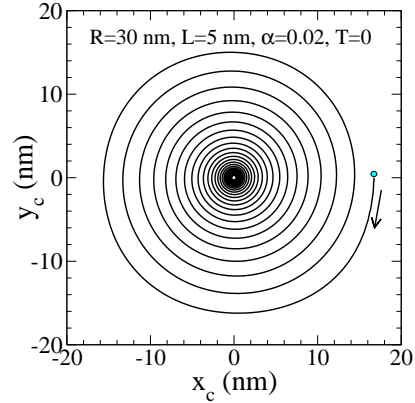}
\caption{\label{vpathZA} Vortex motion with damping, at zero temperature.  This is clockwise motion 
for a vortex with positive ($+\hat{z}$) gyrovector, starting from the dot on the $x$-axis. 
The vortex performs gyrotropic motion of decreasing radius and increasing frequency as it moves 
towards the disk center, ${\bf r}=(0,0)$.}
\end{figure}

The vorticity center position ${\bf r}_v$ is the point around which the in-plane magnetization
components give a divergent curl.  That is, a continuum magnetization field of a vortex 
located at position ${\bf r}_v$, with in-plane angle $\phi({\bf r})$, would be expected to have 
the curl,
\be
\vec\nabla\times \vec\nabla\phi({\bf r}) = 2\pi \hat{z} \delta({\bf r-r}_v) .
\ee
When used on the discrete grid of cells, the vorticity center falls between the four nearest neighbor 
grid cells that have a net $2\pi$ circulation in $\phi$.  However, this discretely defined position
always jumps in increments of the cell size $a$, hence, it cannot be used directly.  Instead, we use
an average position weighted by the squared $m_i^z$ components, of only those cells \textit{near} the 
vorticity center:
\be
{\bf r}_c = \frac{\sum_{|{\bf r}_i-{\bf r}_v|<4\lambda_{\rm ex}}  (m_i^z)^2 \, {\bf r}_i}
{\sum_{|{\bf r}_i-{\bf r}_v|<4\lambda_{\rm ex}} (m_i^z)^2}.
\ee
The ${\bf r}_i$ are the cell positions and the sum is restricted to those cells within four exchange 
lengths of the vorticity center.  The center of the nanodisk is the origin, $(x,y)=(0,0)$.
Including this cutoff in the sums helps to reduce the contributions from other oscillations in 
the system (i.e., spin waves) that are not directly associated with the vortex position.  
By weighting with $(m_i^z)^2$, the position ${\bf r}_c$ is able to change smoothly as the 
vortex moves, especially at $T=0$, in contrast to the discrete vorticity center ${\bf r}_v$.  
It is a reasonable estimate of the mean location of out-of-plane magnetization energy
of the vortex, i.e., close to the vortex core position.  The $m_z$-weighted position ${\bf r}_c$ 
and the vorticity center ${\bf r}_v$ are usually within one lattice constant. 
This measure is supplemented by observing the actual magnetization field when 
there is any doubt about the presence or stability of the vortex.

\begin{figure}
\includegraphics[width=\smallfigwidth,angle=-90]{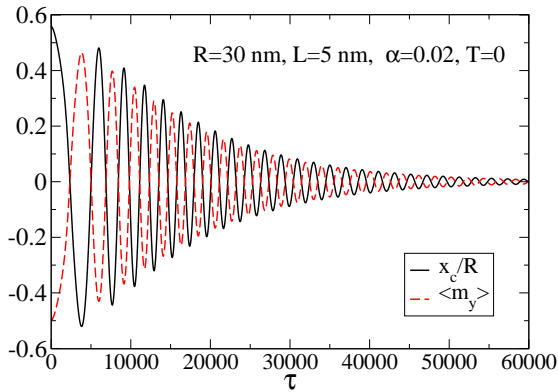}
\caption{\label{xcmyA}  (Color online) For the vortex motion in Figure \ref{vpathZA}, the phase relationship 
between perpendicular components of position and in-plane magnetization.}
\end{figure}

%-------------------------------------------------------------------------------
\subsection{Gyrotropic frequencies in circular disks}
%-------------------------------------------------------------------------------

\begin{figure}
\includegraphics[width=\smallfigwidth,angle=-90]{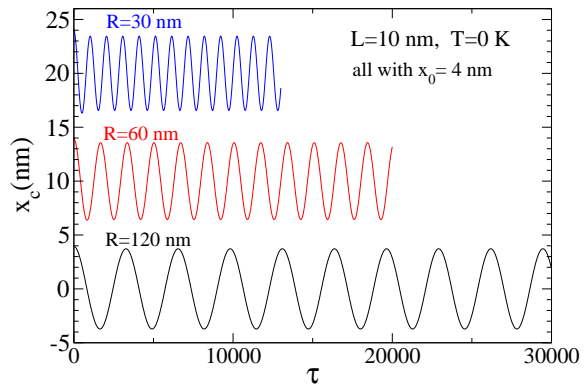}
\caption{\label{xL5} (Color online) Typical motions of the vortex core coordinate $x_c(\tau)$ at zero temperature, 
for circular disks of thickness $L=10$ nm with different radii (shifted vertically from 
$x_c=0$ for clarity).  The damping $\alpha=0.02$ was turned off at time $\tau=1000$.  
Periods were calculated from the energy-conserving motion after $\tau>1000$.  The motion of 
$y_c(\tau)$ is similar but shifted a quarter of a period.}
\end{figure}

Calculations were carried out for circular disks of thickness 5.0 nm, 10 nm and 20 nm 
($L=2.5a, 5a, 10a$,  all with $a=2.0$ nm) for radii 30 nm, 60 nm, 90 nm and 120 nm.  The stability of 
the vortex state is easily checked for a given geometry, by starting from a relaxed vortex
at some radius near half the radius of the disk.  Including a weak damping $\alpha=0.02$,
it is necessary only to run a short simulation of the dynamics and observe whether the
vortex moves in the direction given by the Thiele equation,\cite{Thiele73} Eq.\ (\ref{Vvort}).

For example, with $R=30$ nm, $L=5.0$ nm, a vortex was initially relaxed at a position 
$(x_0,y_0)=(16, 0)$ nm, and then the dynamics was started, including damping $\alpha=0.02$
in the RK4 method.  In this case the vortex is very stable and spirals into the center of the 
disk, see Figures \ref{vpathZA} and \ref{xcmyA}.  The instantaneous vortex displacement on one axis, 
scaled by disk radius, takes approximately the same magnitude as the perpendicular in-plane component 
of $\langle \hat{m} \rangle$, such as $x_c/R$  and $\langle m_y \rangle$ in Figure \ref{xcmyA}.  
Another feature is that the period of rotation becomes less as the vortex moves inward.  The first 
few periods are $\Delta\tau=6000, 3140, 2580,$ but the later revolutions have an average period 
$\tau_G\approx 2020$ (1.51 ns, frequency $f_G=1/\tau_G=0.661$ GHz for Py).

\begin{figure}
\includegraphics[width=\smallfigwidth,angle=-90]{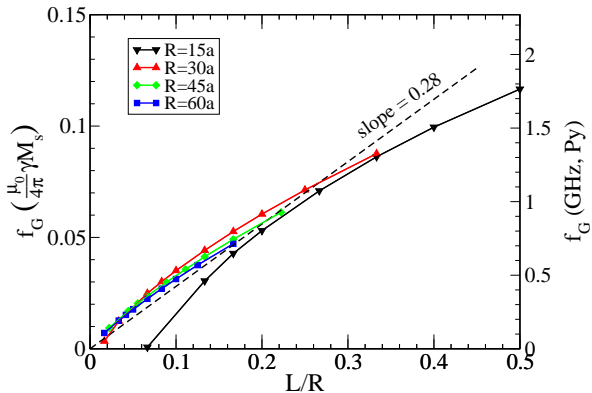}
\caption{\label{f-v-LoR} (Color online) Zero-temperature vortex gyrotropic frequency $f_G$ for various disk 
radii $R$, versus aspect ratio $L/R$. [For Permalloy, $\tfrac{\mu_0}{4\pi} \gamma M_s \approx 15.1$ GHz].  
The computation cell size is $a=2.0$ nm.  The vortex state is unstable below a minimum disk 
thickness, as expected due to the diminished restoring forces from the reduced edge area.
The dashed line shows the result [Eq.\ (\ref{fGlinear})] from using the linear approximation in 
Eq.\ (\ref{kFlinear}) for $k_F$.}
\end{figure}

Other similar dynamics calculations were done at various disk sizes, but turning off the damping 
$\alpha=0.02$ after $\tau=1000$, see Figure \ref{xL5}. 
This initial damped motion is used to remove spin waves that
might be generated when the vortex is initially released, after being relaxed at a desired
starting position. Once the damping is turned off, the dynamics is energy conserving.
Because we are later interested in small movements near the disk center, 
the initial position was taken as $(x_0,y_0)=(2a,0)$, using a lattice constant $a=2.0$ nm. 
These simulations result in very smooth circular motion of the vortex center ${\bf r}_c$ (Fig. \ref{xL5}), 
from which very precise estimates of the gyrotropic period $\tau_G$ were determined by following 
the motion for typically five to ten periods.  The resulting frequencies $f_G$, in units of 
$\tfrac{\mu_0}{4\pi} \gamma M_s$, are shown versus aspect ratio $L/R$ in Figure \ref{f-v-LoR}.  The 
scale is also given there for the parameters of Permalloy, for which 
$\tfrac{\mu_0}{4\pi} \gamma M_s\approx 15.1$ GHz.  One can note the obvious feature, that the 
gyrotropic frequency goes to zero at some minimum thickness needed for vortex stability.

\begin{figure}
\includegraphics[width=\smallfigwidth,angle=-90]{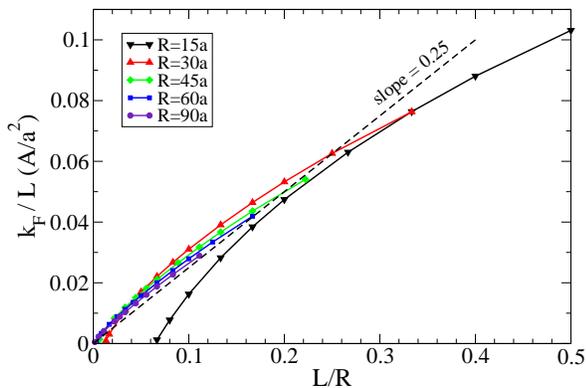}
\caption{\label{kFoL-v-LoR} (Color online)  
Vortex force constant $k_F$ scaled by disk thickness, versus disk aspect ratio. These were 
obtained by assuming a parabolic potential for vortex motion within the disk. The dashed line 
indicates that the slope of this relationship is close to $1/4$ for some range of parameters, 
Eq.\ (\ref{kFlinear}), for disks of adequate thickness. Cell edge is $a=2.0$ nm.}
\end{figure}

%---------------------------------------------------
\subsection{Relation to force constant $k_F$}
%---------------------------------------------------
The vortex restoring force constants $k_F$ were estimated based only on static energy 
considerations.  We compared the total system energy with the displaced 
vortex, $U(x)$, taking $x=2a$, with the energy for the vortex at the disk center, $U(0)$.  
% sentence modified after review:
It is known that the vortex potential is close to parabolic, as long as the vortex
displacement is small compared to the disk radius.\cite{Wysin2010}  The force constant is 
then estimated simply by solving 
\be
\label{Ux}
U(x) = U(0) + \frac{1}{2} k_F \, x^2.
\ee
The energies applied in this equation are those obtained after the vortex is relaxed by the
Lagrange-constrained method.  These calculations are relatively fast because there is no
need to run the dynamics.  The raw force constants were obtained for a wide variety of disk
sizes.  Generally, we find that $k_F$ increases faster than linearly with disk thickness 
$L$ and decreases with disk radius $R$.   

It is expected that the force constant should scale somewhat with the aspect ratio, $L/R$.  
Further, the Thiele equation suggests that the ratio $k_F/L$ is most relevant in determining
$\omega_G$ [see Eq.\ (\ref{wG})].  Therefore, we show $k_F/L$ versus $L/R$ in 
Figure \ref{kFoL-v-LoR}, which presents a relationship somewhat close to linear, with a 
slope near $1/4$.  Thus we can write as a rough  approximation (far enough from the critical 
disk thickness for vortex stability),
\be
\label{kFlinear}
k_F \approx \frac{1}{4} \frac{L^2}{R}\frac{A}{a^2} 
= \frac{\lambda_{\rm ex}^2}{8a^2}\mu_0 M_s^2 \frac{L^2}{R} = 0.878 \mu_0 M_s^2 \frac{L^2}{R}.
\ee
The last form, obtained by applying the definition of exchange length, is preferred because 
the vortex restoring force ultimately is due to the demagnetization fields generated by $M_s$.

\begin{figure}
\includegraphics[width=\smallfigwidth,angle=-90]{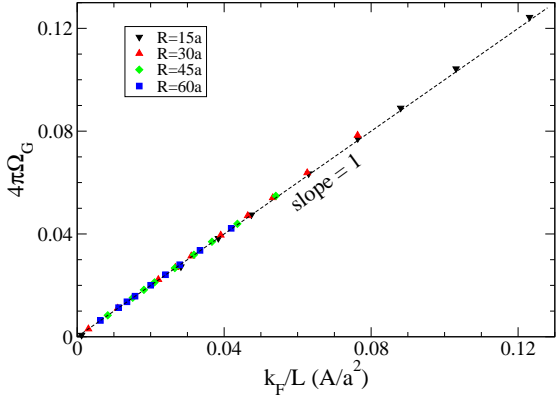}
\caption{\label{nu-v-kFoL} (Color online)
The dimensionless gyrotropic frequencies (found from dynamics) versus force constant scaled 
by disk thickness. The dashed line of unit slope is Eq.\ (\ref{OG}). This verifies the dynamics 
of the Thiele equation, and shows the complete consistency between the static energetics and the 
dynamics. Cell edge is $a=2.0$ nm.}
\end{figure}

One can check whether these force constants are consistent with the gyrotropic frequencies
found in the dynamics.  If the Thiele equation applies to this motion, then the gyrotropic 
frequencies must be linearly proportional to $k_F/L$, [Equations (\ref{wG}) and (\ref{OG})].
Therefore we have plotted the dimensionless frequency $\Omega_G$ versus $k_F/L$ in Figure
\ref{nu-v-kFoL}.  For the wide variety of disk sizes studied, all points in this plot fall on a
single line of unit slope, exactly consistent with the Thiele equation.  This shows that
the calculations of the dynamics over fairly long times (many periods) are completely 
consistent with the force constants found only from static energy considerations.  It
further implies that we can safely use static energy calculations to \textit{predict}
dynamic properties.  This is based on the assumption of an isotropic parabolic potential 
in which the vortex moves.  There may be some limitation to this idea, however, only
because the potential will deviate from parabolic for larger displacements from the
disk center.

These results are consistent with the two-vortices model applied by Guslienko 
\textit{et al.}\cite{Guslienko++02}  With the boundary parameter $\xi=2/3$ and the initial 
susceptibility at small aspect ratio being $\chi(0)^{-1}\approx 9.98 L/R$, their result
(converted to SI units by factor $\tfrac{\mu_0}{4\pi}$) is approximately
\be
k_F = \pi L \frac{\mu_0}{4\pi} M_s^2 \xi^2 \chi(0)^{-1} \approx  1.109 \mu_0 M_s^2 \frac{L^2}{R}.
\ee
Our results have a somewhat weaker potential, which is to be expected because the numerical
simulations allow for a wider range of possible deformations of the vortex structure than is
possible in an analytic approximation.  In addition, our numerical results include the destabilization
of the vortex at sufficiently small $L/R$, hence, it is impossible to fit any straight line for 
$k_F/L$ vs. $L/R$ down to arbitrarily small aspect ratio, see Figure \ref{kFoL-v-LoR}.

We showed above that the gyrotropic frequencies $\nu_G$ are exactly linearly proportional to 
$k_F/L$, hence, this implies that the frequencies also scale close to linearly with $L/R$.
Combining our fit of $k_F$ with relation (\ref{OG}) then shows that roughly, the dimensionless 
angular frequency magnitude is
\be
\Omega_G \approx \frac{1}{16\pi}  \frac{L}{R} \approx  0.0199 \frac{L}{R}.
\ee
In physical units, this is
\be
\label{omG}
\omega_G = \gamma B_0 \Omega_G \approx 0.140\, \gamma \mu_0 M_s \frac{L}{R}  .
\ee
Then the frequency comes out
\be
\label{fGlinear}
f_G = \frac{\omega_G}{2\pi}\approx 0.280 \left(\frac{\mu_0}{4\pi} \gamma M_s\right) \frac{L}{R}.
\ee
The dashed line in Figure \ref{f-v-LoR} shows Eq.\ (\ref{fGlinear}) compared with data from various 
disk sizes.   These frequencies are smaller than those in the rigid vortex model,\cite{Guslienko++01} and 
only slightly smaller than those for the two-vortices model.\cite{Guslienko++02}  
However, this result fits quite well with the experimental data presented in Ref.\ \onlinecite{Guslienko++06}
by also using the higher value for the gyromagnetic ratio, $\gamma = 1.85\times 10^{11}$ s$^{-1}$ T$^{-1}$, 
in conjunction with saturation magnetization still at the value $M_s=860$ kA/m.   The calculation here 
can be considered as that for a more flexible vortex. The magnetization at the edge of the disk 
adjusts itself to try to follow the boundary.  The magnetization can also adjust itself, to a lesser 
extent, in the vortex core region.  These effects lead to lower force constants and therefore
lower gyrotropic frequencies.

These results show that the adapted 2D methods applied here give reliable results, consistent with 
experiment and with the two-vortices analytic calculation of the gyrotropic frequencies.  We note 
that the smaller value of cell constant used here ($a=2.0$ nm) is important for the simulation to 
correctly describe the magnetization dynamics in the vortex core.  Of course, this then imposes a 
limitation on the system size that can be studied.

These results confirm the basic dynamic properties, that the vortex resonance
frequency $\omega_G$ diminishes with increasing dot radius, and increases with increasing
dot thickness.  A wider dot has a weaker spring constant $k_F$ in its potential, 
$U(r)=U(0)+\tfrac{1}{2} k_F r^2$, leading to the reduction of its resonance frequency.  
Similarly, in a thicker dot, the greater area at the edge produces a larger restoring 
force, leading to a higher resonance frequency.

%-------------------------------------------------------------------------------
\section{Thermal effects in vortex dynamics in circular disks}
%-------------------------------------------------------------------------------
In the following part, the effects of thermal fluctuations on the vortex dynamics are 
considered.  We consider two basic situations left to evolve in time via Langevin 
dynamics: (1) A vortex started off-center, and (2) a vortex started at the minimum
energy position, the center of the disk.  In the latter case, the question is whether
thermal fluctuations alone are sufficient to initiate gyrotropic motion.  If so, we can
also study its frequency and range of motion. In all simulations we used cell size
$a=2.0$ nm and damping parameter $\alpha=0.02$ .

\subsection{Vortex initially off-center}

\begin{figure}
\includegraphics[width=\smallfigwidth,angle=-90]{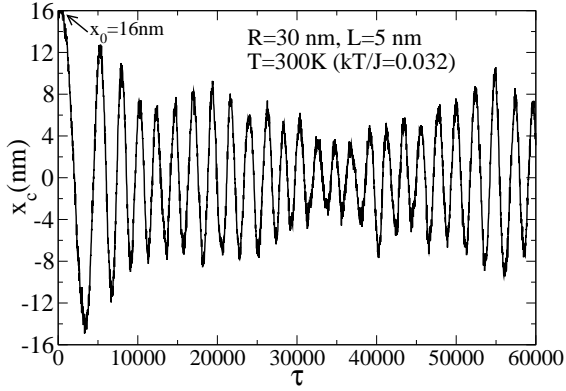}
\includegraphics[width=\smallfigwidth,angle=-90]{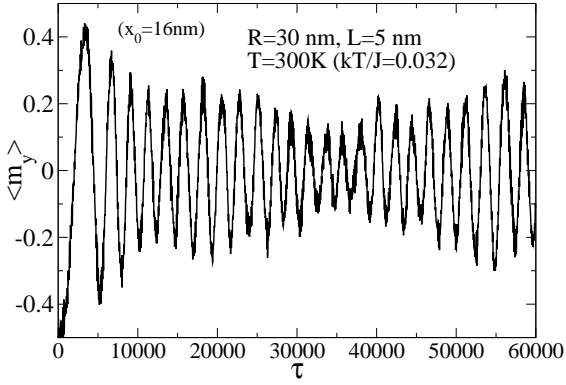}
\caption{\label{xtB}  Vortex motion in Py at room temperature
(300 K), starting from an initial displacement of 16 nm from the disk center.
The $y$-component of average magnetization in the disk is correlated to the $x$-component 
of the vortex position.}
\end{figure}

For the same system used above [$R=30$ nm, $L=5.0$ nm],  
the same initial condition was used, with vortex at $(x_0,y_0)=(16,0)$ nm, but a finite temperature 
corresponding to Permalloy at 300 K was considered. The dynamics was solved now by the H2 scheme.  
The scaled temperature depends on the thickness $L$ of the disk and the exchange stiffness $A$ 
of the material.  The energy unit here is $J=2AL=130$ zJ, while 300 K 
corresponds to $kT=4.14$ zJ, so the scaled temperature is ${\cal T} = kT/J= 0.032$.  The $x$-component
of the vortex position versus time is shown in Figure \ref{xtB}.  In this case, the vortex still 
spirals towards the center of the disk, however,  thermal fluctuations remain present in the motion 
even at time $\tau=60000$ ($\approx 90$ ns), 25 revolutions later.  The range of the motion there 
remains close to $\pm 6$ nm.  The time dependence of $\langle m_y \rangle$ (most closely related to 
$x_c$) is also shown in Figure \ref{xtB}; it also shows an effect persisting at the 25\% level out 
to $\tau=60000$.  Note that at zero temperature, the time-scale for relaxation (Figure \ref{xcmyA}) was
on the order of $\tau \sim 20000$.  This shows that thermal forces apparently are able to maintain the 
gyrotropic motion to very long times.  The average period of the motion is $\tau_G \approx 2278$ 
(1.705 ns, frequency $f=1/\tau_G =0.586$ GHz for Py), showing that the temperature also softened the 
potential experienced by the vortex.

\begin{figure}
\includegraphics[width=\smallfigwidth,angle=-90]{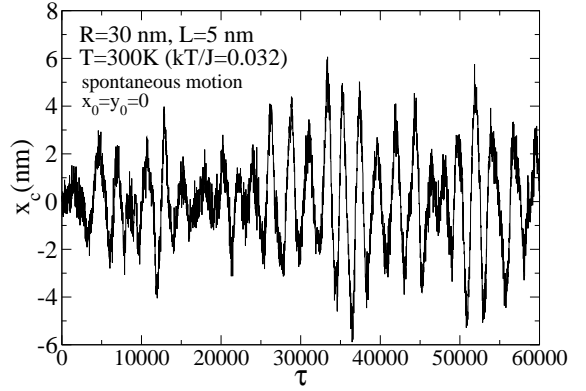}
\caption{\label{xtC}  Spontaneous gyrotropic vortex motion in Py due to 
thermal fluctuations at 300 K, starting from a vortex at the center of the disk.}
\end{figure}

\begin{figure}
\includegraphics[width=\smallfigwidth,angle=-90]{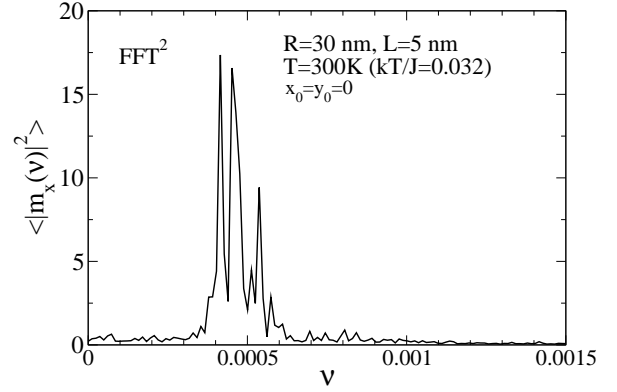}
\caption{\label{mxCfft}  Thermal power spectrum of the in-plane magnetization
fluctuations due to spontaneous gyrotropic vortex motion in Py at 300 K, for 
the motion in Figure \protect\ref{xtC}.}
\end{figure}

\subsection{Vortex initially at disk center}

The same system is used [$R=30$ nm, $L=5.0$ nm], but this time 
the vortex was initiated at the center of the disk, $(x_0,y_0)=(0,0)$.  At zero temperature, such an
initial state is static.  Instead, the dynamics corresponding to Py at 300 K was considered (scaled
temperature ${\cal T}=0.032$).  Any thermal fluctuations can move the vortex core off-center, and if 
that happens, gyrotropic motion can initiate spontaneously.  This indeed happens, as can be seen
in the vortex core position ${\bf r}_c(\tau)$ plotted in Figure \ref{xtC}.  
It needs to be stressed that these vortex motions of the order of $\pm 4$ nm, and magnetization 
fluctuations on the order of $\pm 15$\%, occur without the application of any external magnetic field.  
The motion is sufficiently coherent that it can be followed for dozens of rotations. The gyrotropic 
motion was followed out to
twice the time shown in the plots.  An average over 24 rotations results in a period $\tau_G=2250$,
corresponding to 1.68 ns or a frequency $f=0.594$ GHz.  To verify this, we also show the power
spectrum of the in-plane magnetization oscillations in Figure \ref{mxCfft}.  This was obtained by 
taking time FFTs of $\langle m_x(\tau) \rangle$ of length 256 points at different starting times in 
the data out to $\tau=120000$ and averaging their absolute squares.  The middle peak in Figure 
\ref{mxCfft} falls at dimensionless frequency $\nu\approx 4.52\times 10^{-4}$, corresponding to 
physical frequency $f=\nu/t_0= 0.600$ GHz, consistent with the estimate from counting oscillations.  
There is some structure in the FFT, possibly the beating between three different primary frequencies,
that causes the amplitude of the oscillations to wax and wane.

\begin{figure}
\includegraphics[width=\smallfigwidth,angle=-90]{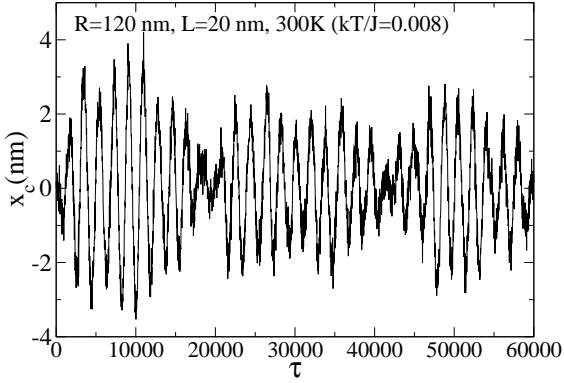}
\caption{\label{xtL} Spontaneous gyrotropic vortex motion, due to thermal fluctuations,
in a 20 nm thick Py disk at 300 K, with the vortex starting at the center of the disk. 
The natural periodic motion executes 32 revolutions in this time sequence, with period 
$\tau_G \approx 1870$.}
\end{figure}

\begin{figure}
\includegraphics[width=\smallfigwidth,angle=-90]{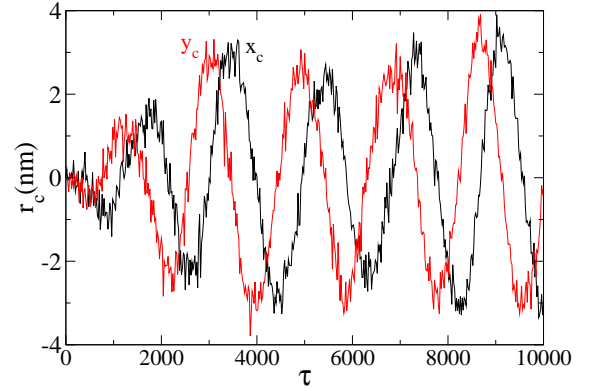}
\includegraphics[width=\smallfigwidth,angle=-90]{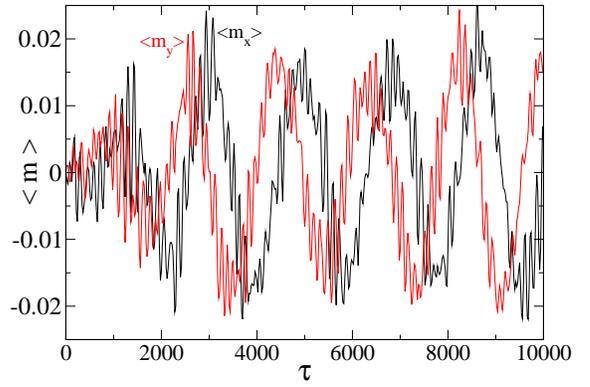}
\caption{\label{xytL}  (Color online) For the spontaneous gyrotropic vortex motion in 
Figure \ref{xtL}, [$R=120$ nm, $L=20$ nm Py disk at 300 K], details of the motion at 
earlier times.  The vortex started at the center of the disk. There is a high-frequency 
spin wave oscillation apparent in the magnetization dynamics, excited together with the
gyrotropic motion.} 
\end{figure}

The spontaneous gyrotropic vortex motion takes place for a wide range of system sizes that 
were tested.  Another example is given for a larger system [$R=120$ nm, $L=20$ nm] in 
Figure \ref{xtL}, where the vortex core displacement is displayed.  An interesting feature is
apparent.  The gyrotropic motion loses its phase coherence at times, leading randomly to brief
intervals of dramatically changed amplitude.  This is only one example; in other time sequences
for other system sizes, this behavior is particularly intermittent and random.  For the same 
simulation, Figure \ref{xytL} also shows both components of vortex core position and both
components of the average in-plane magnetization, zoomed in to show details at earlier times.    
Here one can see the quarter-period phase difference between $x$ and $y$ components for the
vortex position as well as for the magnetization. In addition, the magnetization exhibits a
high-frequency oscillation with a period of about $\Delta\tau\approx 125$ on top of the 
gyrotropic oscillations.  This can be expected to be spin wave excitations that are excited
thermally together with the vortex gyrotropic motion.  

\begin{figure}
\includegraphics[width=\smallfigwidth,angle=-90]{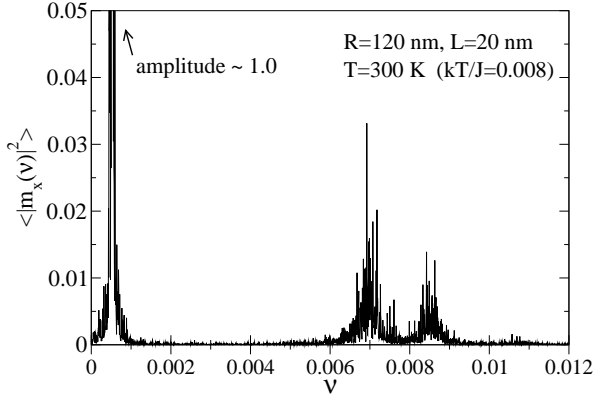}
\caption{\label{mx-doublet}  (Color online) The thermally averaged power spectrum in one component
of the magnetization (squared FFT) for the vortex motion in Figure \ref{xtL}.  
The low frequency gyrotropic mode dominates strongly over a much weaker doublet at high frequency.
For Permalloy parameters ($f={\rm 1336 GHz} \times \nu$), the gyrotropic frequency is 
$f_G = 0.71$ GHz while the components of the doublet lie at $f_1 = 9.3$ GHz and $f_2 = 11.4$ GHz.}
\end{figure}

% paragraph added after review:
To confirm the identity of these spin wave oscillations, we also show in Figure \ref{mx-doublet}
the power spectrum in net magnetization component $m_x$, from a longer simulation out to time
$\tau = 2.5\times 10^5$ .  The vertical scale has been zoomed in to bring out the appearance
of a doublet with frequencies of 9.3 GHz and 11.4 GHz, for Permalloy parameters, while the
gyrotropic frequency is only 0.71 GHz.  A spin wave doublet with azimuthal quantum numbers 
$m=\pm 1$ (wavefunction varying as $\psi \sim e^{im\phi}$ around the disk center) has been
discussed in Ref.\ \onlinecite{Ivanov++10}.  The doublet is predicted to have a 
splitting\cite{Guslienko+2008} of $\Delta f = f_2-f_1 = 3.5 f_G$ and an averaged 
frequency\cite{Zaspel+05}
of $\bar{f} = 1.8 \left( \frac{\mu_0}{4\pi}\gamma M_s \right) \sqrt{\frac{L}{R}}$. For the
situation here, these formulas predict $\Delta f = 2.5$ GHz and $\bar{f} = 11.1$ GHz, while the
observed doublet has $\Delta f = 2.1$ GHz and $\bar{f} = 10.3$ GHz.  Although slightly softer, 
these are of the right orders of magnitude and are consistent with the the theoretical prediction 
for this doublet.  This lowest doublet relates to the presence of spin waves propagating azimuthally
around the disk, in the presence of the vortex.  The splitting can be attributed to the breaking 
of symmetry for the two directions of propagation, due to the presence of the out-of-plane
magnetization  at the vortex core.  Based on these results and results at other disk sizes,
we then note that the primary deviation from a smooth gyrotropic motion is due to the thermal 
excitation of this doublet on top of the vortex magnetization.

% figure added after review:

%------------------------------------------------------------------------
\subsection{Analysis of thermal vortex motion in circular nanodisks}
%------------------------------------------------------------------------
The spontaneous vortex motion at 300 K takes place without the application of
any externally generated magnetic field.  Only the thermal energy is responsible
for the motion.  Indeed, both the frequency and amplitude of this spontaneous
gyrotropic motion is determined directly by the temperature.  Here we give
some analysis and suggest where this motion might be most easily observed
experimentally.

\begin{figure}
\includegraphics[width=\smallfigwidth,angle=-90]{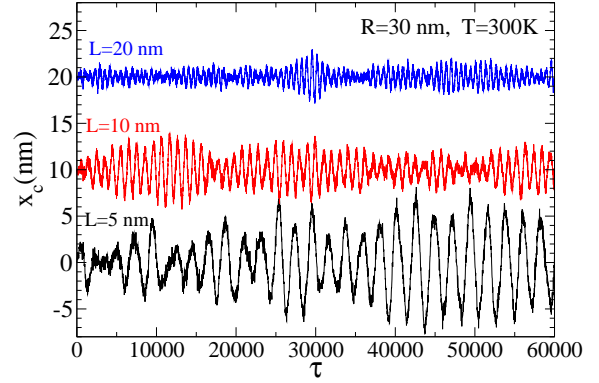}
\caption{\label{xR15T1} (Color online) Typical spontaneous fluctuations of the vortex core 
$x$-coordinate for 30 nm radius Py disks with various thicknesses, at 300 K. The vortex was
initiated at the disk center.  Curves are shifted vertically from $x_c=0$ for clarity.}
\end{figure}

\begin{figure}
\includegraphics[width=\smallfigwidth,angle=-90]{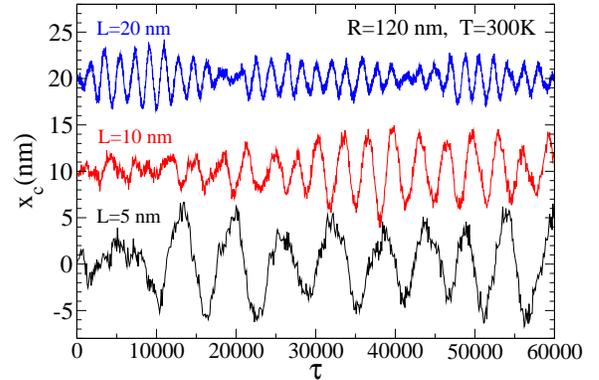}
\caption{\label{xJKL} (Color online) Typical spontaneous fluctuations of the vortex core 
$x$-coordinate for 120 nm radius Py disks with various thicknesses, at 300 K. The vortex was
initiated at the disk center.  Curves are shifted vertically from $x_c=0$ for clarity.}
\end{figure}

For some smaller disks with $R=30$ nm, and for some larger disks, with $R=120$ nm, Figures 
\ref{xR15T1} and \ref{xJKL} exhibit the typical time dependence of the vortex coordinate 
$x_c(\tau)$, for Permalloy systems at 300 K.
The vortex was initially relaxed at the center of the disk ($x=y=0$).  As seen for
the systems studied above, the gyrotropic motion is spontaneous, and furthermore,
takes place at a lower frequency for thinner disks.  In addition, there is a 
dependence of the \textit{amplitude} of the motion on the disk thickness.  The amplitude
is observed to be larger for thinner disks.  Also it is apparent that generally the amplitude 
is larger for the larger radius disks.  This is somewhat difficult to analyze
precisely, due to the limited time sequences that can be obtained during a reasonable
computation time.  However, from knowledge of the force constants $k_F$ and their
dependence on the disk geometry, the RMS range of the vortex core motion can be predicted.

The statistical mechanics of the vortex core position ${\bf X}=(X(t),Y(t))$ and velocity
${\bf V} = \dot{\bf X}$ can be obtained from the effective Hamiltonian associated with the 
Thiele equation.   The Thiele equation is mathematically equivalent to the equation of motion 
for a massless charge $e$ in a uniform magnetic field ${\bf B}$, with $e{\bf B}=-{\bf G}$, 
and also affected by some other force ${\bf F}$.  We can start from a Lagrangian that 
leads to the Thiele equation, using the symmetric gauge for the effective vector potential, and 
including a circularly symmetric parabolic potential (harmonic approximation),
\be
L({\bf X},\dot{\bf X}) = -\tfrac{1}{2} G (X\dot{Y}-Y\dot{X}) - \tfrac{1}{2} k_F (X^2+Y^2) . 
\ee
The first term on the RHS is equivalent to $e{\bf V}\cdot {\bf A}$, with vector potential
${\bf A}= \tfrac{1}{2}{\bf B}\times {\bf X}$ in a magnetic problem; there is no usual
kinetic energy term like $\tfrac{1}{2}m {\bf V}^2$, because the intrinsic mass is considered
zero here.  Only the $z$-component of the gyrovector is present, $G \equiv G_z=2\pi pq m_0 \gamma^{-1}$.
Then the components of the Thiele equation are recovered from the Euler-Lagrange variations,
\bn
\frac{\partial L}{\partial X}-\frac{d}{dt}\frac{\partial L}{\partial \dot{X}}= -k_F X -  G \dot{Y} = 0, \\
\frac{\partial L}{\partial Y}-\frac{d}{dt}\frac{\partial L}{\partial \dot{Y}}= -k_F Y +  G \dot{X} = 0. 
\en
The Lagrangian is written equivalently as
\be
L({\bf X},{\bf V}) = -\tfrac{1}{2} ( {\bf G} \times {\bf X} ) \cdot {\bf V} - \tfrac{1}{2} k_F {\bf X}^2.
\ee
This leads to the canonical momentum,
\bn
\label{Pdef}
{\bf P} = \frac{\partial L}{\partial \bf V} = -\tfrac{1}{2} {\bf G} \times {\bf X} = (\tfrac{G}{2}Y, -\tfrac{G}{2}X).
\en
This allows the transformation to the collective coordinate Hamiltonian, $H({\bf X},{\bf P})$.
Following the usual prescription, we have
\be
\label{HX}
H({\bf X},{\bf P})={\bf P}\cdot \dot{\bf X} - L = \frac{1}{2} k_F {\bf X}^2
=  \frac{1}{2} k_F \left(X^2+Y^2\right).
\ee
Note that the derivation of the Hamiltonian does not depend on the choice of the gauge for
the gyrovector (i.e., for its effective magnetic field).  In Ref.\ \cite{Ivanov+2010}, it is
shown that the Landau gauge leads to the same result for $H$, but where $P=GY$ is found to
be the momentum conjugate to $X$. 

Technically this is all that is needed to analyze the statistics of the vortex position. By being
purely potential energy, however, this Hamiltonian needs careful treatment.  Its variation via the 
Hamiltonian equations of motion does not lead back to the correct dynamics, i.e., it does not give the 
Thiele equation.  One can see that the
difficulty is due to the fact that the position and canonical momentum coordinates are redundant,
since $P_x = \tfrac{1}{2}G Y$ and $P_y = - \tfrac{1}{2}G X$.   Even so, all of these should be 
considered linearly independent mechanical coordinates, and all should appear in $H$ to give the 
correct dynamics (gyrotropic motion does not conserve ${\bf X}$ nor ${\bf P}$, so both should appear 
in $H$). For that to work out, $H$ must be expressed so that there are both potential and kinetic 
energy terms. (A similar care is needed even in the Landau gauge, where $GY$ must be identified by 
and replaced as the momentum $P$ conjugate to $X$.) We can split out half of the potential 
energy and redefine it in terms of ${\bf P}^2$ as a kinetic energy,
\be
\label{HXP}
H({\bf X},{\bf P})= 
\tfrac{1}{4} k_F {\bf X}^2 + \tfrac{1}{4} k_F \left(\frac{2 \bf P}{G}\right)^2 .
\ee
One can easily demonstrate that the correct dynamic equations result only by allocating exactly 
half of the energy as kinetic energy and half as potential energy.  This then leads to the Hamilton 
dynamic equations for oscillations along the two perpendicular axes.  For example, along $x$ there is
\bn
\label{dotX}
\dot{X} = \frac{\partial H}{\partial P_x} = \frac{2k_F P_x}{G^2} ,  \\
\label{dotP}
\dot{P}_x = -\frac{\partial H}{\partial X} = -\tfrac{1}{2}k_F X . 
\en
These give a second order equation for simple harmonic motion (SHO), 
\be
\label{SHO}
\ddot{X} = -\frac{k_F^2}{G^2} X.
\ee
The other variations with respect to $Y$ and $P_y$ lead to the same dynamics for $Y$.  However, note
that the Thiele equation is recovered from these dynamics only by including the connection (\ref{Pdef})
that defines the canonical momentum in terms of the position.

It is clear that the Hamiltonian (\ref{HXP}) is the same as that for a two-dimensional simple harmonic
oscillator with coordinate ${\bf X}$ and momentum ${\bf P}$.  For that oscillator, the effective 
spring constant is $k_{\rm SHO}=\tfrac{1}{2}k_F$, and the corresponding effective mass is 
$m_{\rm SHO} = \frac{G^2}{2k_F}$.   It is interesting to see that these lead back to the natural 
frequency of gyrotropic motion [or see Eq.\ (\ref{SHO})],
\be
\omega_G = \omega_{\rm SHO} = \sqrt{\frac{k_{\rm SHO}}{m_{\rm SHO}}} = \frac{k_F}{G}.
\ee
Of course, as $G$ is proportional to the disk thickness via the factor $m_0=LM_s$, 
and $k_F$ depends on both $R$ and $L$, then this contains the various geometrical effects,
especially those associated with the vortex force constant. 

\begin{figure}
\includegraphics[width=\smallfigwidth,angle=-90]{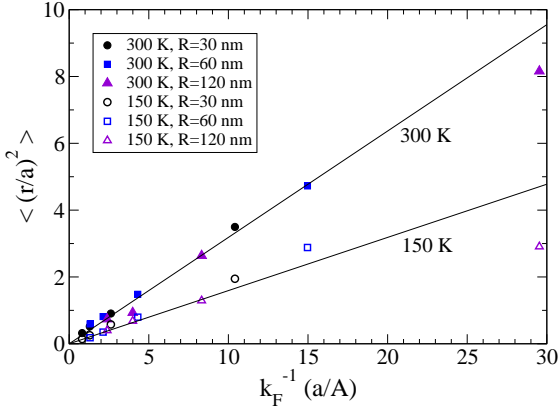}
\caption{\label{r2-v-kFi} (Color online) 
Average squared displacement of the vortex core from the disk center, versus reciprocal
force constant.  The points come from simulations out to time $\tau=2.5\times 10^5$; the
solid lines are the predictions from the equipartition theorem, Eq.\ \ref{r2}, using 
the parameters for Py.}
\end{figure}

In consideration of the classical statistical mechanics, the important fact here is that 
the Hamiltonian (\ref{HX}) has a dynamics due to only two coordinates ($X,Y$) appearing 
quadratically.  Although the dynamic equations for $\dot{X}$ and $\dot{Y}$ must come
from the Hamiltonian (\ref{HXP}) of the equivalent 2D SHO, the phase space of the Thiele dynamics 
is more restricted, due to relation (\ref{Pdef}) between ${\bf P}$ and ${\bf X}$.  This
forces the Thiele phase space to be only two dimensional; this does not depend on the
choice of the gauge. As an example of that reduction of
the phase space, elliptical motions are present for the 2D SHO, while the zero-temperature 
Thiele dynamics has only circular orbits.  As we are considering thermal 
equilibrium, each \textit{independent} quadratic coordinate receives an average thermal energy of 
$\tfrac{1}{2}k T$.  This gives the connection needed to predict the average RMS vortex displacement 
from the disk center.  Specifically, for each vortex core coordinate,
\be
\langle \tfrac{1}{2} k_F X^2 \rangle = \langle \tfrac{1}{2} k_F Y^2 \rangle = \tfrac{1}{2} k T.
\ee
Then the average squared displacement of the vortex from the disk center should be
\be
\label{r2}
\langle r^2 \rangle = \langle X^2+Y^2 \rangle = r_{\rm rms}^2 = \frac{2 k T}{k_F}.
\ee
These show that the average thermal energy in the vortex motion must be
\be
\langle H({\bf X},{\bf P}) \rangle = k T.
\ee
Therefore, we can check that these relations actually hold in the simulations. The  average 
squared displacement should be proportional to the reciprocal of the force constant, with
the same proportionality factor (twice the temperature) when disks of different geometries 
are considered.  Some results for the average squared displacements
versus reciprocal force constant in different geometries are given in Figure \ref{r2-v-kFi}.   
The results depend on the behavior of the force constant with disk geometry, showing the importance 
of  static calculations for understanding the statistical dynamics behavior.  The simulation data
have a general trend consistent with Eq.\ \ref{r2}, but there are large fluctuations due to
the finite time sequences used, which is more of a problem for the systems with small $k_F$. 

\begin{figure}
\includegraphics[width=\smallfigwidth,angle=-90]{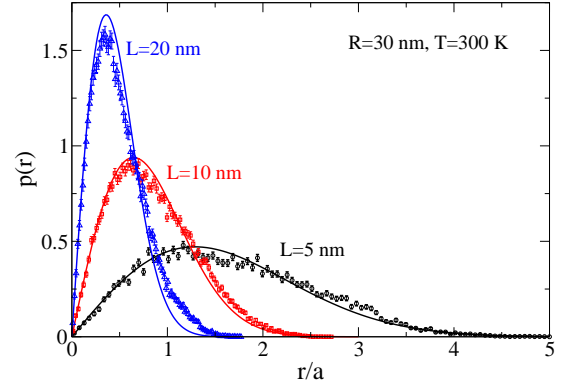}
\includegraphics[width=\smallfigwidth,angle=-90]{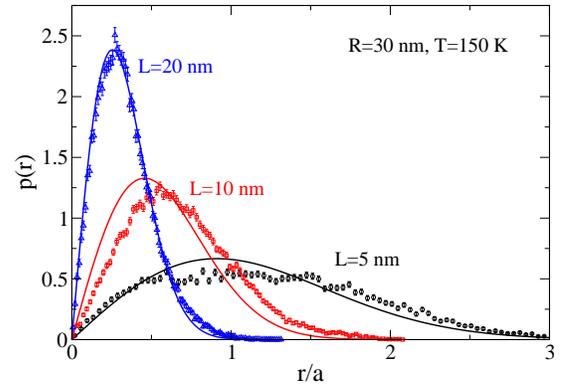}
\caption{\label{pr15} (Color online) Probability distributions in Py disks of radius 30 nm at 
temperatures 300 K and 150 K, for the radial position $r$ of the vortex, measured from the disk 
center, in units of the cell size, $a=2$ nm.  Solid curves are the theoretical expression (\ref{pr}) 
based on a Boltzmann distribution using the static force constants; points are from simulations out 
to time $\tau=2.5\times 10^5$ .}
\end{figure}

\begin{figure}
\includegraphics[width=\smallfigwidth,angle=-90]{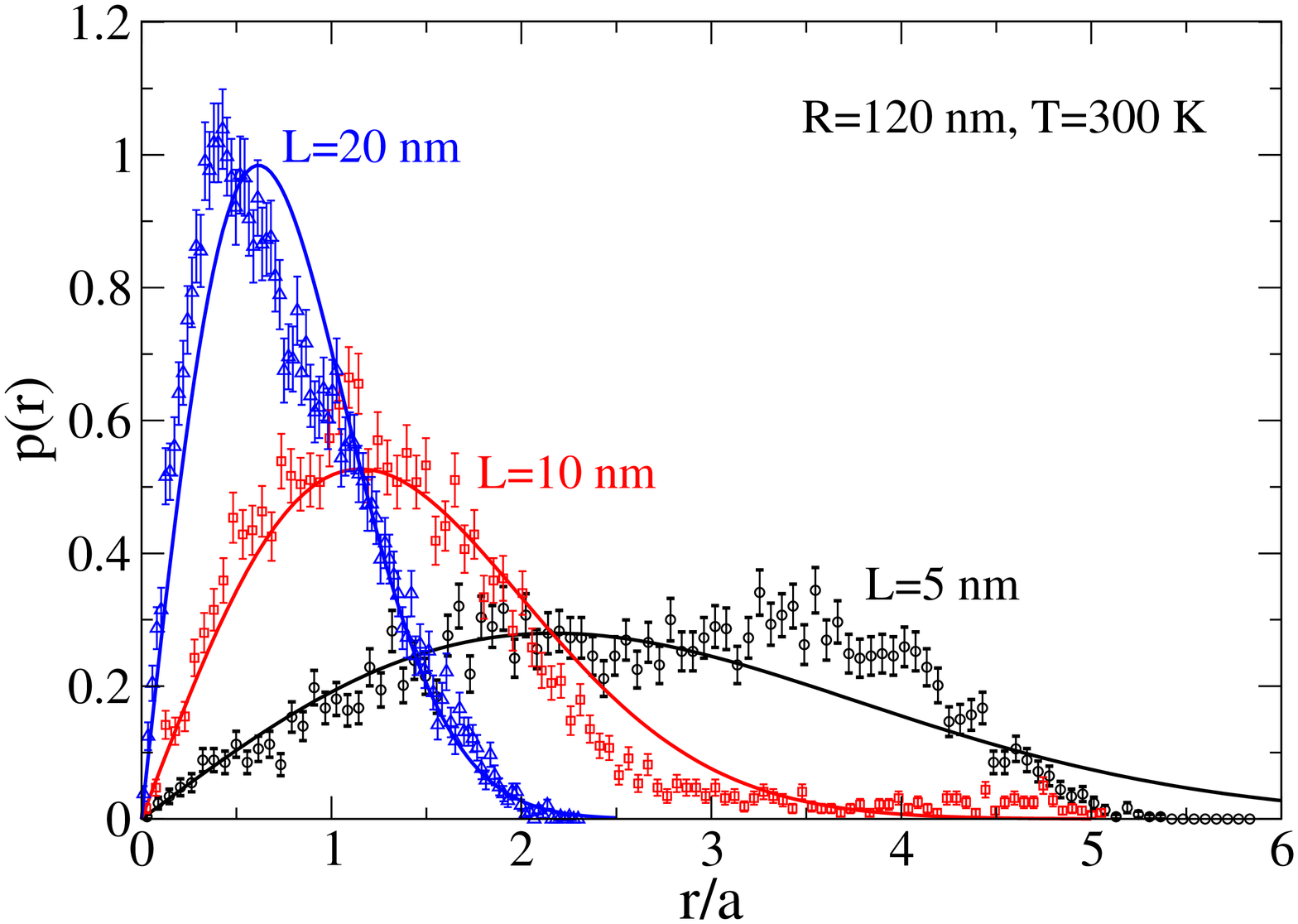}
\includegraphics[width=\smallfigwidth,angle=-90]{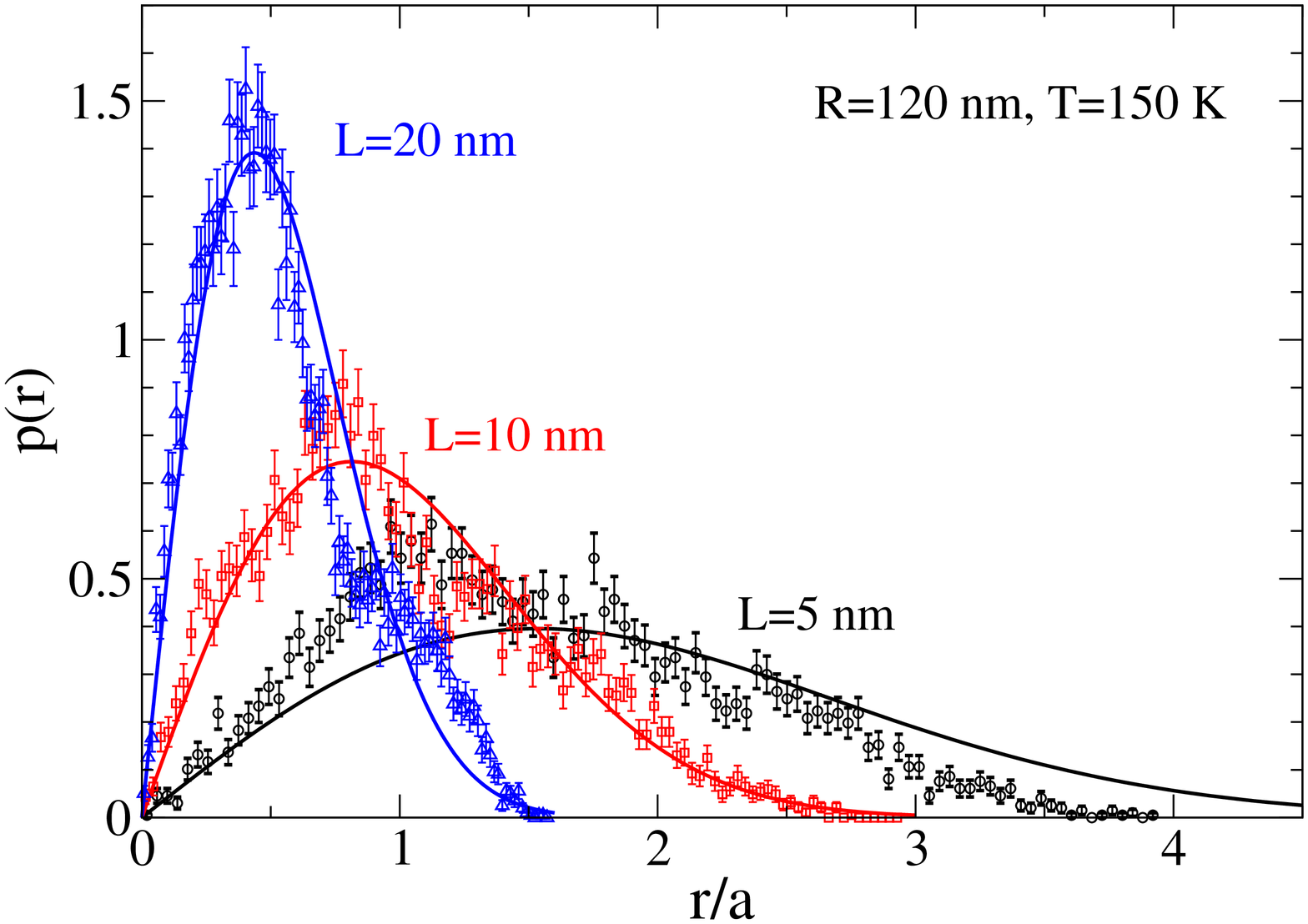}
\caption{\label{pr60} (Color online) Probability distributions for vortex radial position in Py 
disks of radius 120 nm, as explained in Figure \ref{pr15}.}
\end{figure}

We can further substantiate the statistical behavior of the vortex core, by calculating the
probability distribution $p(r)$ of its distance $r=\sqrt{X^2+Y^2}$ from the disk center. 
Assuming that its position is governed by Boltzmann statistics for Hamiltonian (\ref{HX}), the 
normalized distribution from $p(r) dr \propto 2\pi r \, dr \, e^{-\beta H}$ is predicted to be
\be
\label{pr}
p(r) = \beta k_F r \, e^{-\frac{1}{2}\beta k_F r^2} ,
\ee
where $\beta=(kT)^{-1}$ is the inverse temperature.  This distribution also has some particular
distinctive points that are relatively easy to check.  For instance, the distribution has a
peak at the point of maximum probability, at the radius
\be
r_{\rm max} = \sqrt{\frac{kT}{k_F}} = \frac{r_{\rm rms}}{\sqrt{2}}.
\ee
In addition, the value of the function at this point is 
\be
p_{\rm max} = p(r_{\rm max}) = \frac{e^{-1/2}}{r_{\rm max}}.
\ee
We have found that the vortex core position satisfies this distribution reasonably well,
while the vortex is undergoing the spontaneously generated gyrotropic motion.  
There is a certain difficulty to verify this, because very long time sequences
(we used final time $\tau = 250000$) are needed so that many gyrotropic revolutions
are performed. During the motion, at times there are rather large fluctuations in
the amplitude of the motion.  The motion varies between time intervals of smooth
gyrotropic motion of large amplitude and other time intervals where the motion
seems to be impeded, and is of much smaller amplitude.  Even so, we were able to 
take these long sequences and produce histograms of the vortex radial position
to compare with the predicted probability distribution.  An example for $R=30$ nm  
is given in Figure \ref{pr15}. The temperatures are defined here by applying
the material parameters for Permalloy (that is, 300 K corresponds to $kT=0.1592 Aa$,
where the exchange stiffness for Py is $A=13$ pJ/m and cell size $a=2.0$ nm was used in 
all simulations).  The data (points) are compared with the prediction of equation (\ref{pr}) 
(solid curves), for different disk thicknesses.  For these smaller systems, the agreement is
quite good between the simulations and the theoretical expression, Eq.\ (\ref{pr}).

The distributions were also found in simulations for larger radius, see Figure \ref{pr60}
for the distribution at $R=120$ nm. In this case, the errors are considerably greater.  This 
is due primarily to the larger gyrotropic period.   Over the sampling time interval to 
$\tau=2.5 \times 10^5$, there are less periods being sampled.  The system has a somewhat erratic 
behavior, in that the orbital radius of the vortex motion seems to switch suddenly between 
different values, as already mentioned.  As a result, at this system size a greater time interval 
is needed to obtain a sample that could be considered in thermal equilibrium, with well defined averages.

For thinner disks, the number of
revolutions in the given time interval is lesser, which means the thinner disks may
also require longer time sequences to give the same relative errors.  Of course, the thinner 
(thicker) disks have a weaker (stronger) force constant, leading to the greater 
(lesser) amplitude spontaneous motions.  This is clearly exhibited in the probability 
distributions.  Although these aspects may be difficult to verify experimentally, the results 
do indeed point to much stronger spontaneous gyrotropic fluctuations for very thin magnetic disks.  
In the cases where these motions were of greater amplitude, there may start to appear deviations 
from the distribution in (\ref{pr}), simply because the larger amplitude vortex motions cause 
the vortex to move out of the region where the potential is parabolic.

%-------------------------------------------------------------------
\section{Discussion and conclusions}
%-------------------------------------------------------------------
The calculations here  give a precise description of the
magnetostatics and dynamics for thin-film nanomagnets, especially in the
situations where a single vortex is present.  The continuum problem for some
finite thickness $L$ has been mapped onto an equivalent 2D problem, i.e.,
the modified micromagnetics adapted here.  For high aspect ratios, $L\ll 2R$,
the shape anisotropy is very strong, and this 2D system is a very good approximation
of the full 3D problem, because it leads to the physical situation where the magnetization 
has little dependence on $z$ and is predominately planar, except in the vortex core.  

At zero temperature, we have been able to test this approach and compare with the 
predictions for vortex gyrotropic motion based on the Thiele equation.  This comparison 
is made possible here because the vortex force constants $k_F$ can be calculated from
the energetics of a vortex with a \textit{constrained position}.  The application
of the Lagrange undetermined multipliers technique\cite{Wysin2010} for enforcing a desired 
static vortex position ${\bf X}$ has been essential in the determination 
of $k_F$.  In addition, that relaxation procedure also is of great utility for 
initiating a vortex at some radius while removing most of the initial spin wave like 
oscillations that would otherwise be generated when the time dynamics is started.
As a result, we have been able to determine the zero temperature gyrotropic 
frequencies for the motion of the vortex core, ${\bf X}(t)$, to fairly high precision.  
The confirmation of the applicability of the Thiele equation to the $T=0$ dynamics of 
vortex velocity ${\bf V}$ is impressive, as demonstrated in the straight line fit for 
gyrotropic frequency $\omega_G$ versus scaled force constant $k_F/L$ in Figure \ref{nu-v-kFoL}.   
This shows the complete consistency between the \textit{statics} calculations of the force 
constants and the \textit{dynamics} calculations of the frequencies, when interpreted
via the Thiele equation. 

At larger disk radii, the gyrotropic frequency is found to be close to linear in the 
aspect ratio, $L/R$, see Eq.\ \ref{omG}.  The frequencies are also close 
to those found in the two-vortices model and micromagnetics calculations carried 
out in Ref.\ \onlinecite{Guslienko++02}.  The differences from those results
may be due to the fact that we have used the cell parameter $a$ half of what was 
used in Ref.\ \onlinecite{Guslienko++02}.  This is important, because the cell 
parameter should be sufficiently less than the exchange length for results to be 
reliable.  Otherwise, if $a$ is too large, the details of the energetics and dynamics 
in the vortex core cannot be correctly represented.

At $T>0$, the Langevin dynamics shows some surprising behavior that was
reported earlier in Ref.\ \onlinecite{Machado+11}, even when
the vortex is initiated at the center of a nanodisk.  The thermal fluctuations
are indeed sufficiently strong to produce a spontaneous motion of the vortex
core, without the application of any external field, which is not a simple random walk.  
Instead, the gyrotropic nature of the motion is still present, and in fact, persistent 
vortex rotation is the dominant feature of the motion.   The thermal fluctuations 
can be viewed as a perturbation on top of the gyrotropic motion, however, it is the 
temperature that determines the expected squared radius of the orbit.  The orbital 
radius is very well described from the statistical mechanics of the vortex collective 
coordinate Hamiltonian (\ref{HX}), that possesses only the potential energy associated 
with the vortex force constant.

Integrations of the dynamics over very long times (equivalent to hundreds of vortex
revolutions) shows that the statistics of the vortex position follows the simple
Boltzmann distribution in Eq.\ (\ref{pr}).  The average squared vortex displacement
from the origin, $r_{\rm rms}^2$, scales linearly in the temperature divided \textit{only} 
by the force constant $k_F$.   This is in contrast to the vortex gyrotropic frequencies,
which depend on $k_F/L$.  Thus, the results for force constant indirectly
predict the expected position fluctuations.  However,  very long time sequences
are needed to see this average behavior;  over some short time intervals there
can be large variations in the instantaneous vortex orbital radius.  The
largest spontaneous vortex position fluctuations will be possible in thin dots
of larger radius, where the force constants are weakest.  Even so, this is a
small effect (RMS radii on the order of several nanometers), and it may be difficult 
to observe experimentally.  As an example based only on the calculated force constants,
a magnetic dot of radius $R=180$ nm and thickness $L=20$ nm has $k_F \approx 0.29 A/a$.  
For Py at 300 K, this gives the estimate $r_{\rm rms}\approx 2.1$ nm.  If the thickness 
is reduced to 10 nm, then $k_F\approx 0.080 A/a$ and the RMS orbital radius increases to 
$r_{\rm rms}\approx 4.0$ nm.  Even though these are rather small,  the distributions 
$p(r)$ are rather wide and therefore at times one can expect even larger vortex 
gyrotropic oscillations.

Finally we note that the thermal distribution of the vortex rotational velocity is 
connected to the radial distribution $p(r)$, because the Hamilton equations (\ref{dotX}) 
imply
\be
{\bf V} = \vec\omega_G \times {\bf X}, \quad \vec\omega_G = -\frac{k_F}{G}\hat{z}.
\ee
Thus, we can transform magnitudes with $V=\omega_G r$. 
Then the RMS rotational velocity is
\be
V_{\rm rms} = |\, \omega_G|\, r_{\rm rms} = \frac{\sqrt{2k_F \, kT}}{G},
\ee
which varies proportional to $\sqrt{k_F}/L$.  This is connected to a Boltzmann
distribution for the probability $f(V) \, dV$ of vortex speed $V$ in some interval
of width $dV$,  where
\be
f(V) = \frac{p(V/\omega_G)}{\omega_G} = \beta m_G V e^{-\frac{1}{2}\beta m_G V^2}.
\ee
This involves a gyrotropic effective mass $m_G$, 
\be
\quad m_G\equiv \frac{G^2}{k_F} \approx \frac{(2\pi)^2}{0.878} \frac{R}{\mu_0 \gamma^2},
\ee
determined both by the vortex force constant and by the disk thickness contained in
the definition of $G$.  For small aspect ratio, however, the thickness cancels 
and this mass is proportional to the disk radius alone.  At $R=100$ nm, the mass is
about $1.2 \times 10^{-22}$ kg, independent of the material.  Although $f(V)$ has a 
mathematical form identical to that for $p(r)$, it leads to another interesting 
interpretation of the vortex dynamics in equilibrium.

%---------------------------------------------------------------------------
\section*{Acknowledgments}
%---------------------------------------------------------------------------
G.\ M.\ Wysin acknowledges the financial support of FAPEMIG grant BPV-00046-11 
and the hospitality of Universidade Federal de Vi\c cosa, Minas Gerais, Brazil, 
and of Universidade Federal de Santa Catarina, Florian\'opolis, Brazil, where 
this work was carried out during sabbatical leave. W.\ Figueiredo acknowledges
the financial support of CNPq (Brazil).

\end{document}